\documentclass[PAPER,atlasdraft=true,cernpreprint,texlive=2018,UKenglish,LANGEDIT=true,LANGSHOW=false,texmf]{atlasdoc}
 
\usepackage{atlaspackage}
\usepackage{comment}
\usepackage{tabularx}
\usepackage{atlasbiblatex}
 
\usepackage{atlasphysics}
 
\graphicspath{{logos/}{figures/}}
 
\usepackage{ANA-HIGG-2018-52-PAPER-defs}
\usepackage{atlasjetetmiss}
\usepackage{multirow}
\usepackage{rotating}
\usepackage{alphalph}
\usepackage{float}
\usepackage{dcolumn}
 
\usepackage{booktabs}
\renewcommand{\arraystretch}{1.2}

\AtlasTitle{Measurement of the associated production of a Higgs boson decaying into $b$-quarks with a vector boson at high transverse momentum in \pp\ collisions at  \sqsthirt with the ATLAS detector}

\AtlasJournalRef{Phys. Lett. B 816 (2021) 136204}
\AtlasDOI{10.1016/j.physletb.2021.136204}
 
\AtlasAbstract{
 
The associated production of a Higgs boson with a $W$ or $Z$ boson
decaying into leptons and where the Higgs boson decays to a $\bbbar$
pair is measured in the high vector-boson transverse momentum regime,
above 250~\GeV, with the ATLAS detector. The analysed data,
corresponding to an integrated luminosity of \intlumi, were collected
in proton--proton collisions at the Large Hadron Collider between
2015 and 2018 at a centre-of-mass energy of \sqsthirt. The measured
signal strength, defined as the ratio of the measured signal yield to
that predicted by the Standard Model, is $0.72 ^{+0.39}_{-0.36}$
corresponding to an observed (expected) significance of 2.1 (2.7)
standard deviations. Cross-sections of associated production of a
Higgs boson decaying into $b$ quark pairs with a $W$ or $Z$ gauge
boson, decaying into leptons, are measured in two exclusive
vector boson transverse momentum regions, 250--400~\GeV\ and above 400~\GeV, and
interpreted as constraints on anomalous couplings in the framework of
a Standard Model effective field theory.

}
 
\author{The ATLAS Collaboration}
 
\AtlasRefCode{HIGG-2018-52}

\hypersetup{pdftitle={ATLAS document},pdfauthor={The ATLAS Collaboration}}
 
\begin{document}
 
\maketitle

\section{Introduction}
\label{sec:intro}
Since the discovery of the Higgs boson
($H$)~\autocite{Englert:1964et,Higgs:1964ia,Higgs:1964pj,Guralnik:1964eu}
with a mass of around 125~\GeV~\cite{HIGG-2014-14} by the ATLAS
and CMS Collaborations~\autocite{HIGG-2012-27,CMS-HIG-12-028} in 2012,
the analysis of proton--proton ($pp$) collision data at centre-of-mass
energies of 7~\TeV, 8~\TeV\ and 13~\TeV\ delivered by the Large Hadron
Collider (LHC)~\autocite{Evans:2008zzb} has led to precise
measurements of the main production cross-sections and decay rates of
the Higgs boson, as well as measurements of its mass and its spin and
parity properties. In particular, the observation of the decay of the
Higgs boson into $b$-quark pairs provided direct evidence for the
Yukawa coupling of the Higgs boson to down-type
quarks~\cite{Aaboud:2018zhk,Sirunyan:2018kst}.  Finally, a combination
of 13~\TeV\ results searching for the Higgs boson produced in
association with a leptonically decaying $W$ or $Z$ boson established
the observation of this production process~\cite{Aaboud:2018zhk}.
A first cross-section measurement as a
function of the vector-boson transverse momentum was also carried out
by the ATLAS Collaboration~\cite{HIGG-2018-50}.

The previous ATLAS analyses~\cite{Aaboud:2018zhk,HIGG-2018-50} in this
channel were mainly sensitive to vector bosons with transverse
momentum ($\pt$) in the range of approximately 100--300~\GeV. These
analyses considered a pair of jets with radius parameter of $R= 0.4$,
referred to as small-radius (\smallR) jets, to reconstruct the Higgs
boson. For higher Higgs boson transverse momenta, the decay products
can become close enough that they cannot be reconstructed with
two \smallR\ jets.  To explore this `boosted' regime, the Higgs boson
is reconstructed as a single \largeR\ jet with
$R=1.0$~\cite{Butterworth:2008iy}.
This high-\pt\ regime is particularly interesting due to
its sensitivity to physics beyond the Standard
Model~\cite{Mimasu:2015nqa}.

This Letter presents a measurement of cross-sections for the
associated production of a high transverse momentum Higgs boson that
decays into a \bbbar pair with a leptonically decaying $W$ or $Z$
boson. The analysis uses \pp\ collision data recorded between 2015 and
2018 by the ATLAS detector~\cite{PERF-2007-01} during Run~2 at the
LHC. This dataset corresponds to an integrated luminosity
of \intlumi. Events are selected in 0-, 1- and 2-lepton channels,
based on the number of reconstructed charged leptons, $\ell$
(electrons or muons), in the final state to explore the $ZH \to \nu\nu
b\bar{b}$, $WH \to \ell \nu b \bar{b}$ and $ZH \to \ell\ell b\bar{b}$
signatures, respectively. The Higgs boson is reconstructed as a
single \largeR\ jet and the $b$-quarks from its decay as a pair of
jets, reconstructed with a \pt-dependent radius parameter, associated
with the \largeR\ jet and identified as containing a $b$-hadron.
 
The analysis using \smallR\ jets and focusing on slightly lower Higgs
boson transverse momentum regions was recently updated with the complete Run~2
dataset~\cite{CompanionPaper}.
The \largeR\ jet analysis
significantly overlaps with the \smallR\ jets analysis. The two
results can therefore not be straightforwardly combined.

The dominant background processes after the event selection correspond
to the production of \vjets, where $V$ refers to either a $W$ or $Z$
boson, \ttbar, single-top and dibosons.  The signal is extracted from
a combined profile likelihood fit to the \largeR jet mass, using
several signal and control regions. The yield of diboson production
$VZ$ with $Z \rightarrow b\bar{b}$ is also measured using the same fit
and provides a validation of the analysis. The cross-section
measurements are performed within the simplified template
cross-section (STXS) framework~\cite{deFlorian:2016spz,
Badger:2016bpw}. These measurements are then used to constrain
anomalous couplings in a Standard Model effective field theory
(SMEFT)~\cite{Contino:2013kra}.


\section{ATLAS detector}
\label{sec:detector}
\newcommand{\AtlasCoordFootnote}{ATLAS uses a right-handed coordinate
system with its origin at the nominal interaction point in the
centre of the detector. The positive $x$-axis is defined by the
direction from the interaction point to the centre of the LHC ring,
with the positive $y$-axis pointing upwards, while the beam
direction defines the $z$-axis. Cylindrical coordinates $(r,\phi)$
are used in the transverse plane, $\phi$ being the azimuthal angle
around the $z$-axis. The pseudorapidity $\eta$ is defined in terms
of the polar angle $\theta$ by $\eta=-\ln\tan(\theta/2)$. The
angular distance is defined as $\DeltaR \equiv \sqrt{(\Delta \eta)^2
+ (\Delta \phi)^2}$. Rapidity is defined as $y$ = $0.5 \ln[(E +
p_z)/(E-p_z)]$ where $E$ denotes the energy and $p_z$ is the
component of the momentum along the beam direction.}
 
The ATLAS detector~\cite{PERF-2007-01} at the LHC is a multipurpose particle detector with
a forward--backward symmetric cylindrical geometry and a near $4\pi$
coverage in solid angle.\footnote{\AtlasCoordFootnote} It consists of
an inner detector (ID) for tracking surrounded by a thin
superconducting solenoid providing a \SI{2}{\tesla} axial magnetic
field, electromagnetic and hadronic calorimeters, and a muon
spectrometer.  The ID covers the pseudorapidity range $|\eta| < 2.5$.
It consists of silicon pixel, silicon microstrip, and transition
radiation tracking detectors.  An inner pixel layer, the insertable
B-layer~\cite{ATLAS-TDR-2010-19, Abbott:2018ikt}, was added at a mean
radius of 3.3~cm during the long shutdown period between Run~1 and
Run~2 of the LHC.  Lead/liquid-argon (LAr) sampling calorimeters
provide electromagnetic (EM) energy measurements with high granularity
($|\eta| < 3.2$). The hadronic calorimeter uses a
steel/scintillator-tile sampling detector in the central
pseudorapidity range ($|\eta| < 1.7$) and a copper/LAr detector in the
region $1.5 < |\eta| < 3.2$. The forward regions~($3.2< |\eta| <4.9$)
are instrumented with copper/LAr and tungsten/LAr calorimeter modules
optimised for electromagnetic and hadronic measurements, respectively.
A muon spectrometer with an air-core toroid magnet system surrounds
the calorimeters.  Three layers of high-precision tracking chambers
provide coverage in the range $|\eta|<2.7$, while dedicated fast
chambers allow muon triggering in the region $|\eta|<2.4$.  The ATLAS
trigger system consists of a hardware-based first-level trigger
followed by a software-based high-level trigger~\cite{TRIG-2016-01}.
 
\section{Data and Monte Carlo simulation}
\label{sec:dataMC}
The data were collected in \pp collisions at \sqsthirt\ during Run~2 of the LHC.
The data sample corresponds to an integrated luminosity of
139~\ifb~after requiring that all detector subsystems were operating
normally and recording high-quality data~\cite{Aad:2019hmz}. The
uncertainty in the combined 2015--2018 integrated luminosity is
1.7\% \cite{ATLAS-CONF-2019-021}, obtained using the LUCID-2
detector \cite{Avoni:2018iuv} for the primary luminosity measurements.
Collision events considered for this analysis were recorded with a
combination of triggers selecting events with high missing transverse
momentum or with a high-\pt\ lepton, depending on the analysis
channel. More details of the trigger selection are given in
Section~\ref{sec:selection}.
 
Monte Carlo (MC) simulated event samples processed with the ATLAS detector
simulation~\cite{SOFT-2010-01} based
on \GEANT4~\cite{Agostinelli:2002hh} are used to model the signal and
background contributions, except for the multijet production, whose
contribution is estimated with data-driven techniques as detailed in
Section~\ref{sec:background}. A summary of all the signal and background processes
with the corresponding generators used for the nominal samples is
shown in Table~\ref{tab:samples}. All simulated processes are
normalised using the most precise theoretical predictions currently
available of their cross-sections. In addition to the hard scatter,
each event was overlaid with additional \pp collisions (pile-up)
generated with \PYTHIAV{8.1}~\cite{Sjostrand:2007gs} using the ATLAS
A3 set of tuned parameters~\cite{ATL-PHYS-PUB-2016-017} and the
NNPDF23LO~\cite{Ball:2012cx} parton distribution function (PDF)
set. Simulated events were then reconstructed with the same algorithms
as those applied to data and are weighted to match the pile-up
distribution observed in the data.
 
For the signal events, the \textsc{AZNLO}~\cite{Aad:2014xaa} model of
parton showers and the underlying event (UE) was used. For the
top-quark pair and single-top-quark production processes, the UE
model was taken from the ATLAS A14~\cite{ATL-PHYS-PUB-2014-021} set
of tuned \PYTHIAV{8.1}~\cite{Sjostrand:2007gs} parameters and for
the other backgrounds the default \textsc{
Sherpa}~\cite{Gleisberg:2008ta,Bothmann:2019yzt,Cascioli:2011va,Gleisberg:2008fv}
tune set was used. For all samples of simulated events, except for
those generated using \textsc{Sherpa}, the \textsc{EvtGen v1.2.0}
program~\cite{Lange:2001uf} was used to describe the decays of
bottom and charm hadrons. The nominal PDF set used for $W/Z$+jets
and diboson processes was NNPDF3.0NNLO~\cite{Ball:2014uwa}
while for the top-quark pair and single-top production the
NNPDF3.0NLO~\cite{Ball:2014uwa} set was used. Samples produced with
alternative generators which are used to estimate modelling
systematic uncertainties are described in
Section~\ref{sec:systematics}.
 
\begin{table}[tb!]
\begin{center}{\fontsize{7}{7.9}\selectfont
\caption
{
Signal and background processes with the corresponding generators used for the nominal samples.
If not specified, the order of the cross-section
calculation refers to the expansion in the strong coupling constant
($\alphas$).  $(\star)$ The
events were generated using the first PDF in the NNPDF3.0NLO set and
subsequently reweighted to the PDF4LHC15NLO
set~\cite{Butterworth:2015oua} using the internal algorithm in
\textsc{Powheg-Box v2}.  $(\dagger)$ The NNLO(QCD)+NLO(EW)
cross-section calculation for the $pp \to ZH$ process already
includes the $gg\to ZH$ contribution.  The $qq\to ZH$ process is
normalised using the cross-section for the $pp \to ZH$ process,
after subtracting the $gg\to ZH$ contribution. An additional scale
factor is applied to the $qq \rightarrow VH$ processes as a function
of the transverse momentum of the vector boson, to account for
electroweak (EW) corrections at NLO\@. This makes use of the $VH$
differential cross-section computed with
\textsc{Hawk}~\cite{Denner:2011id,Denner:2014cla}.
\protect\label{tab:samples}}
\begin{tabular}{llllll}
\toprule
Process & ME generator & ME PDF &PS and & UE model & Cross-section \hspace{2.5cm}\\
& & & Hadronisation & tune & order\\
\toprule
\multicolumn{6}{l}{Signal ($m_H = 125$~GeV and $b\bar{b}$ branching fraction set to 58\%)} \\
\midrule
$qq\to WH\to\ell\nu b\bar{b}$  &\textsc{Powheg-Box v2}~\cite{Alioli:2010xd} +&
NNPDF3.0NLO$^{(\star)}$~\cite{Ball:2014uwa} &\PYTHIAV{8.212}~\cite{Sjostrand:2014zea} & AZNLO~\cite{Aad:2014xaa} & NNLO(QCD)+  \\
&   \textsc{GoSam}~\cite{Cullen:2011ac} + \textsc{MiNLO}~\cite{Hamilton:2012np,Luisoni:2013kna}  & & & &NLO(EW)~\cite{Ciccolini:2003jy,Brein:2003wg,Ferrera:2011bk,Brein:2011vx,Ferrera:2013yga,Ferrera:2014lca,Campbell:2016jau} \\
$qq\to ZH\to\nu\nu b\bar{b}/\ell\ell b\bar{b}$ &\textsc{Powheg-Box v2} + & NNPDF3.0NLO$^{(\star)}$ &\PYTHIAV{8.212} & AZNLO & NNLO(QCD)$^{(\dagger)}$+  \\
&   \textsc{GoSam} + \textsc{MiNLO}  & & & &NLO(EW) \\
$gg  \to ZH\to\nu\nu b\bar{b}/\ell\ell b\bar{b}$ &\textsc{Powheg-Box v2} & NNPDF3.0NLO$^{(\star)}$ &\PYTHIAV{8.212} & AZNLO & NLO+  \\
&  & & & &NLL~\cite{Altenkamp:2012sx,Hespel:2015zea,Harlander:2014wda,Harlander:2013mla,Brein:2012ne}\\
\midrule
\multicolumn{6}{l}{Top quark ($m_t = 172.5$~\GeV)}  \\
\midrule
$t\bar{t}$ &\textsc{Powheg-Box v2}~\cite{Alioli:2010xd,Frixione:2007nw} &  NNPDF3.0NLO &\PYTHIAV{8.230} &A14~\cite{ATL-PHYS-PUB-2014-021}& NNLO+NNLL~\cite{Czakon:2011xx} \\
$s$-channel &\textsc{Powheg-Box v2}~\cite{Alioli:2009je,Alioli:2010xd} & NNPDF3.0NLO &\PYTHIAV{8.230} &A14 & NLO~\cite{Kidonakis:2010tc} \\
$t$-channel &\textsc{Powheg-Box v2}~\cite{Alioli:2009je,Alioli:2010xd} & NNPDF3.0NLO &\PYTHIAV{8.230} &A14& NLO~\cite{Kidonakis:2011wy} \\
$Wt$ &\textsc{Powheg-Box v2}~\cite{Re:2010bp,Alioli:2010xd} & NNPDF3.0NLO &\PYTHIAV{8.230} &A14& Approximate NNLO~\cite{Kidonakis:2010ux} \\
\midrule
\multicolumn{6}{l}{Vector boson + jets} \\
\midrule
$W\to\ell\nu$ & \textsc{Sherpa 2.2.1}~\cite{Gleisberg:2008ta,Bothmann:2019yzt,Cascioli:2011va,Gleisberg:2008fv} & NNPDF3.0NNLO &  \textsc{Sherpa 2.2.1}~\cite{Schumann:2007mg,Hoeche:2012yf} & Default & NNLO~\cite{Catani:2009sm}  \\
$Z/\gamma^{*}\to\ell\ell$ & \textsc{Sherpa 2.2.1} & NNPDF3.0NNLO &  \textsc{Sherpa 2.2.1} & Default & NNLO    \\
$Z\to\nu\nu$ & \textsc{Sherpa 2.2.1} & NNPDF3.0NNLO &  \textsc{Sherpa 2.2.1} & Default & NNLO    \\
\midrule
\multicolumn{6}{l}{Diboson} \\
\midrule
$qq \to WW$ & \textsc{Sherpa 2.2.1} & NNPDF3.0NNLO &  \textsc{Sherpa 2.2.1} & Default & NLO  \\
$qq \to WZ$ & \textsc{Sherpa 2.2.1} & NNPDF3.0NNLO &  \textsc{Sherpa 2.2.1} & Default & NLO  \\
$qq \to ZZ$ & \textsc{Sherpa 2.2.1} & NNPDF3.0NNLO &  \textsc{Sherpa 2.2.1} & Default & NLO  \\
$gg \to VV$ & \textsc{Sherpa 2.2.2} & NNPDF3.0NNLO &  \textsc{Sherpa 2.2.2} & Default & NLO  \\
\bottomrule
\end{tabular}}
\end{center}
\end{table}

All $qq$-initiated signal processes were simulated with up to one
additional parton at next-to-leading-order (NLO) accuracy in QCD using
the  \POWHEGBOXV{v2}~\cite{Alioli:2010xd}
and the \textsc{GoSam}~\cite{Cullen:2011ac} generator with the MiNLO
(Multiscale Improved NLO)~\cite{Hamilton:2012np,Luisoni:2013kna}
procedure applied, interfaced to
\PYTHIAV{8.212}
for the simulation of the
parton shower (PS), UE and multiple parton interactions. The
$gg \rightarrow ZH$ contribution was simulated at leading order (LO)
in QCD with {\textsc Powheg-Box}~v2. The
$gg \rightarrow ZH$ cross-section process was calculated at NLO in QCD
including soft gluon resummation up to next-to-leading logarithms
(NLL)~\cite{Altenkamp:2012sx,Hespel:2015zea,Harlander:2014wda,Harlander:2013mla,Brein:2012ne}. Signal
MC events were generated using the NNPDF3.0NLO PDF set and
subsequently reweighted to the PDF4LHC15NLO PDF
set~\cite{Butterworth:2015oua}. The total inclusive cross-sections for
all signal processes ($WH$ and $ZH$) were calculated at
next-to-next-to-leading-order (NNLO) QCD and NLO electroweak
(EW)~\cite{Ciccolini:2003jy,Brein:2003wg,Ferrera:2011bk,Brein:2011vx,Ferrera:2013yga,Ferrera:2014lca,Campbell:2016jau}
accuracy, including photon-induced contributions calculated with
\textsc{Hawk}~\cite{Denner:2011id,Denner:2014cla}.
 
The nominal top-quark pair production generator
was \textsc{Powheg-Box}~v2 with real and virtual corrections at NLO
accuracy in QCD and interfaced to
\textsc{Pythia}~8.230 for the parton showering. The nominal top-quark pair
production cross-section is from a resummed NNLO and
next-to-next-to-leading logarithm (NNLL) prediction~\cite{Czakon:2011xx}.
 
Single top-quark production was also generated
with \textsc{Powheg-Box}~v2 interfaced to \textsc{Pythia}~8.230. The
nominal cross-section normalisations for the single top-quark
production $s$- and $t$- processes were estimated from resummed
calculations at NLO, while for the $Wt$ process approximate NNLO was
used~\cite{Kidonakis:2010tc,Kidonakis:2011wy,Kidonakis:2010ux}. At
higher orders in QCD, the definition of the $Wt$ process
can correspond to leading-order top-quark pair production
processes. To account for these ambiguities and related
interference effects when generating the processes separately,
the diagram removal (DR)
subtraction scheme was used~\cite{Frixione:2008yi}.
 
The nominal $W/Z$+jets background samples used \textsc{Sherpa
2.2.1}~\cite{Bothmann:2019yzt,Cascioli:2011va,Gleisberg:2008fv} for
the matrix element (ME) and parton shower with virtual corrections at NLO accuracy for up to
two additional jets and at LO for up to four additional jets
using \textsc{OpenLoops}~\cite{Gleisberg:2008ta,Cascioli:2011va,Gleisberg:2008fv}.
In these samples, the simulation of the emission of hard partons
matched with a parton shower was based on the Catani--Seymour
subtraction
term~\cite{Gleisberg:2008ta,Cascioli:2011va,Gleisberg:2008fv} and
the multi-parton ME was merged with the parton shower
using an improved \textsc{ckkw} matching procedure extended to NLO
accuracy using the \textsc{MEPS@NLO}
prescription~\cite{Hoeche:2012yf}. The nominal normalisation of this
background was obtained from an NNLO fixed-order
estimate~\cite{Catani:2009sm}.
 
The diboson nominal samples were generated using \textsc{Sherpa 2.2.1}
for the dominant $qq$-initiated processes for which zero or
one additional parton was calculated at NLO in the ME,
while two or three additional partons were included at LO in QCD. The
subdominant $gg$-initiated processes were generated
with \textsc{Sherpa 2.2.2}. For these samples, zero or one additional parton was
calculated at LO in the ME. These generators also provided the nominal
normalisation for this process.

 
\section{Object reconstruction}
\label{sec:reconstruction}

Of all the reconstructed
$pp$ collision vertices with at least two reconstructed trajectories
of charged particles in the ID (tracks) with $\pt>0.5$~\GeV,
the hard-scattering primary vertex is selected as the
one with the highest sum of squared transverse momenta of associated
tracks~\cite{ATL-PHYS-PUB-2015-026}.

Leptons are used for event categorisation as described in
Section~\ref{sec:selection}.  Electrons are reconstructed from tracks
in the ID associated with topological clusters of energy
depositions in the calorimeter~\cite{ATL-PHYS-PUB-2017-022,
Aad:2019tso}. The identification criteria closely follow those
described in Ref.~\cite{Aaboud:2018zhk}. \textit{Baseline} electrons
are required to have \pt\ $>$ 7~\GeV\ and $\abseta<2.47$,
to be isolated from other tracks and energy deposit clusters, to meet
loose likelihood selection criteria based on shower shapes and to
satisfy $|d_0/\sigma(d_0)| < 5$ and $|z_0\sin\theta|<0.5$~mm,
where $d_0$ and $z_0$ are the transverse and longitudinal impact
parameters defined relative to the primary vertex
position\footnote{For the computation of the impact parameters, the
beam line is used to approximate the primary vertex position in the
transverse plane.} and $\sigma(d_0)$ is the $d_0$ uncertainty.
\textit{Signal} electrons are a subset of the baseline electron set and are selected
using a tighter likelihood requirement, which also includes tracking
and track--cluster matching variables, and using a tighter
calorimeter-based isolation criterion.

Muon candidates are identified by matching
ID tracks to full tracks or track segments reconstructed in the muon
spectrometer within the inner detector coverage and using only information
from the muon spectrometer outside of that coverage. Muons are required to have $\pt > 7$~\GeV\ and
$|\eta|<2.7$ and to have $|d_0/\sigma(d_0)| < 3$ and $|z_0\sin\theta|<0.5$~mm. Two muon
categories are used in the analysis: \textit{baseline} muons are selected
using the `loose' identification criterion of Ref.~\cite{PERF-2015-10} and a loose track isolation;
\textit{signal} muons are required to have $|\eta|<2.5$, to satisfy the `medium' identification criterion~\cite{PERF-2015-10}
and a tighter track-based isolation criterion.

The low-threshold (7~\GeV) \textit{baseline} leptons are used to
define the three main channels requiring exactly zero, one and two
leptons. The latter 1- and 2-lepton channels further require at least
one \textit{signal} lepton, with identification and isolation
requirements chosen to optimise the suppression of the multijet
background. Signal leptons must have a $\pt>27$~\GeV\ (except in
the 1-lepton muon sub-channel where a $\pt>25$~\GeV\ is used).
 
Calorimeter jets are reconstructed from noise-suppressed topological
clusters (topoclusters) of calorimeter energy
depositions~\cite{PERF-2014-07}, using the \antikt\
algorithm~\cite{Cacciari:2008gp} with radius parameter $R=1.0$
(\largeR\ jets) or $R=0.4$ (\smallR\ jets) implemented in \textsc{
FastJet}~\cite{Fastjet}. \SmallR\ jets are built from topoclusters
calibrated at the electromagnetic scale~\cite{PERF-2016-04},
while \largeR\ jets are built from topoclusters calibrated at the
local hadronic scale~\cite{PERF-2014-07}. \LargeR\ jets are groomed
using trimming~\cite{Krohn:2009th,PERF-2012-02} to improve the jet
mass resolution and its stability with respect to \pileup by
discarding the softer components of jets that originate from
initial-state radiation, \pileup interactions, or the underlying
event. This is done by reclustering the constituents of the initial
\largeR\ jet, using the \kt\
algorithm~\cite{Catani:1993hr,Ellis:1993tq}, into subjets with radius
parameter $\Rsub=0.2$ and removing any subjet that has a \pT\ less
than 5\% of the parent jet \pT. The \largeR\ jet mass $m_\textrm{J}$
is computed using tracking and calorimeter
information~\cite{Aaboud:2018kfi}.  A dedicated MC-based calibration,
similar to the procedure used in Ref.~\cite{Aaboud:2018kfi}, is
applied to correct the \pt\ and mass of the trimmed jets to the
particle level. \LargeR\ jets are required to have $\pt > 250$~\GeV,
$m_\textrm{J}>50$~\GeV\ and $|\eta| < 2.0$, the last due to tracking
acceptance.
 
\SmallR\ jets are used in building the missing transverse momentum
and event categorisation.  They are
calibrated with a series of simulation-based corrections and $in$ $situ$
techniques, including corrections to account for pile-up energy
entering the jet area, as described in Ref.~\cite{PERF-2016-04}. They
are required to have $\pT > 30$~\GeV\ and $|\eta| < 4.5$. To reduce
the number of \smallR\ jets originating from pile-up interactions, \smallR\ jets
are required to pass the jet vertex tagger
(JVT)~\cite{PERF-2014-03} requirement if they are in the range
$\pT < 120$~\GeV\ and $|\eta| < 2.5$ due to tracking acceptance.
 
Track-jets formed from charged-particle tracks are used to reconstruct
a candidate two-body \Hbb decay within the
\largeR jet.  Track-jets are built with the \antikt\
algorithm with a variable radius (VR) \pt-dependent parameter, from
tracks reconstructed in the inner detector with $\pt > 0.5$~\GeV\ and
$|\eta| < 2.5$~\cite{Krohn:2009zg, ATL-PHYS-PUB-2017-010,
Aad:2019uoz}. VR track-jets have an effective jet radius
$R_\textrm{eff}$ proportional to the inverse of the jet \pt\ in the
jet finding procedure: $R_\textrm{eff}(\pt) = \rho/\pt$, where the
$\rho$-parameter is set to 30~\GeV. There are two additional
parameters, $R_\textrm{min}$ and $R_\textrm{max}$, used to set the
minimum and maximum cut-offs on the jet radius, and these are set to
0.02 and 0.4, respectively. Only VR track-jets with $\pt>10$~\GeV,
$|\eta| < 2.5$ and with at least two constituents are considered~\cite{PERF-2016-05}.
VR track-jets are matched to the \largeR calorimeter jets via
ghost-association~\cite{Cacciari_2008}. Track-jets not associated with \largeR jets are also used in the
analysis for event categorisation as described in
Section~\ref{sec:selection}.

The `truth' flavour labelling of track-jets in simulation
is done by geometrically matching the jet to `truth' hadrons, using
`truth' information from the generator's event record. If a $b$-hadron
with \pT\ above 5~\GeV\ is found within $\Delta R = 0.3$ of the
direction of the track-jet, the track-jet is labelled as a $b$-jet.
If the $b$-hadron is matched to more than one
track-jet, only the closest track-jet is labelled as a $b$-jet. If no
$b$-hadron is found, the procedure is repeated first for $c$-hadrons
to label $c$-jets and then for $\tau$-leptons to label $\tau$-jets.  As is the case for
defining a $b$-jet, the labelling is also exclusive for $c$- and $\tau$-jets. A
jet for which no such matching can be made is labelled as a light-flavour jet.

To identify track-jets containing $b$-hadron decay products,
track-jets are tagged using the multivariate algorithm MV2c10,
which exploits the presence of large-impact-parameter tracks, the
topological decay chain reconstruction and the displaced vertices from
$b$-hadron decays~\cite{Aad:2019aic,ATL-PHYS-PUB-2017-013}.
The MV2c10 algorithm is configured to achieve an average efficiency of 70\% for
tagging jets labelled as $b$-jets in an MC sample of $\ttbar$ events. This requirement
has corresponding rejection factors of 9 and 304 for jets labelled as
$c$-jets and light-flavour jets, respectively, in simulated $\ttbar$ events.
The tagging efficiencies
per jet flavour are corrected in the simulation to match those measured in
data~\cite{PERF-2016-05,ATLAS-CONF-2018-006,ATLAS-CONF-2018-001}.
 
Two additional corrections are applied to the \largeR jets to improve
the scale and the resolution of their energy and mass
measurements. First, to account for semileptonic decays of the
$b$-hadrons, the four-momentum of the closest reconstructed
non-isolated muon candidate within $\Delta R=\mathrm{min}(0.4,
0.04+10$~\GeV$/\pt^{\textrm{muon}})$ of a track-jet matched to
the \largeR\ jet by ghost association is added to the
calorimeter-based component of the \largeR\ jet
four-momentum while its expected calorimeter energy deposits are removed~\cite{Aad:2019uoz}. This is known as the muon-in-jet
correction. Non-isolated muons satisfy the `medium' identification
criterion~\cite{PERF-2015-10}, but no isolation or impact parameter
criteria are applied. Second, in the 2-lepton channel only, a
per-event likelihood uses the full reconstruction of the event
kinematics to improve the estimate of the energy of the $b$-jets~\cite{HIGG-2016-29}. The
kinematic fit constrains the $\ell^+ \ell^- b\overline{b}$ system and the
additional \smallR\ jets in the event to be balanced in the transverse
plane and the dilepton system to the $Z$ boson mass, by scaling
the four-momentum of the objects in the event including the \largeR\ jet,
additional \smallR\ jets and leptons within their detector response resolutions. The \largeR\ jet mass is then
scaled by the ratio of the energies after and before the
correction. For the event selection detailed in
Section~\ref{sec:selection}, the \largeR\ jet mass resolution improves
by 5\% to 10\% after the first correction (depending on the lepton
channel), while the second correction brings an additional improvement
in the 2-lepton channel of up to 40\%.

The presence of neutrinos in the $WH \to \ell \nu b \bar{b}$ and
$ZH \to \nu\nu b\bar{b}$ signatures can be inferred from a momentum
imbalance in the transverse plane. The missing transverse momentum
$\vecmet$ is reconstructed as the negative vector sum of the momenta
of leptons and \smallR\ jets in the event plus a
`soft term' built from additional tracks associated with the primary
vertex~\autocite{PERF-2016-07}. \SmallR\ jets used for the $\vecmet$ reconstruction
are required to have $\pT > 20$~\GeV. The magnitude of $\vecmet$ is referred
to as \met. To suppress non-collision and multijet backgrounds in the
0-lepton channel, an additional track-based missing transverse momentum
estimator, \vecmpt, is built independently as the negative vector sum
of the transverse momenta of all tracks from the primary vertex.

An overlap removal procedure is applied to avoid double-counting
between reconstructed leptons~\cite{Aaboud:2018zhk}, including
hadronically decaying $\tau$-leptons~\cite{PERF-2014-06}, and \smallR\
jets~\cite{HIGG-2016-29}.
 
\section{Event selection}
\label{sec:selection}
Events are categorised into the 0-, 1- and 2-lepton channels
depending on the number of selected electrons and muons to target the
$ZH \to vv b\bar{b}$, $WH \to \ell \nu b \bar{b}$ and $ZH \to \ell\ell
b\bar{b}$ signatures, respectively.
 
The 0-lepton selection is applied to events selected with an \met\
trigger with thresholds varying from 70 to
110~\GeV\ depending on the data-taking period to cope with increasing
trigger rates at higher instantaneous luminosities. In the 1-lepton channel,
single-electron events are required to be triggered by at least one of
several unprescaled single-electron triggers. The lowest $E_\textrm{T}$ threshold of
these unprescaled triggers varied with time from 24 to 26~\GeV.
Events in the single-muon channel were triggered using the same \met\
trigger as used in the 0-lepton channel. Given that muons do not enter
in the online \met\ calculation and that uninstrumented regions affect
the coverage of the muon spectrometer, the \met\ triggers translate into
a requirement on the transverse momentum of the lepton and neutrino pair, $\pT^{\ell\nu}$, which is more efficient in the
analysis phase space than the single-muon triggers.
In the 2-lepton channel, the same trigger strategy as in the 1-lepton
channel is adopted. The dielectron selection is applied to
events triggered by at least one of the un-prescaled single-electron
triggers. The dimuon selection is applied to events triggered by
an \met\ trigger.  All triggers used in this analysis are fully
efficient for the events selected using the requirements described
below.

In all three channels, events are required to contain at least
one \largeR jet with $\pt>250$~\GeV\ and $|\eta|<2.0$. To select the
Higgs boson candidate, the leading \pt\ \largeR\ jet is chosen, at
least two VR track-jets are required to be matched to it by
ghost-association, and the two leading ones are required to be
$b$-tagged.  This jet is referred to as the `Higgs-jet candidate' in
the following. To avoid the ambiguous cases of concentric jets, events
where the $b$-tagged VR track-jets overlap with other VR track-jets,
satisfying $\Delta R / R_\textrm{s}<1$ (where $\Delta R$ corresponds
to the distance among any pair of VR track-jets and $R_\textrm{s}$
corresponds to the smaller radius of the considered pair), are
removed.
The reconstructed transverse
momentum \ptv\ of the vector boson corresponds to \met\ in the
0-lepton channel, to the magnitude of the vector sum of $\vecmet$ and
the charged-lepton transverse momentum in the 1-lepton channel, and to
the transverse momentum of the 2-lepton system in the 2-lepton
channel. The \ptv\ is required to be above 250~\GeV\ in all three
channels. The event selection is detailed in
Table~\ref{tab:EventSelection}, with further explanations provided
below for the non-straightforward selection criteria.
 
\begin{table}
{\fontsize{9}{9.9}\selectfont
\centering
\caption{Event selection requirements for the boosted $VH, H\to$ \bbbar analysis channels and sub-channels.}
\setlength{\extrarowheight}{2pt}
\begin{tabular}{l | c|c|c|c|c|c}
\toprule
Selection & \multicolumn{2}{c|}{0 lepton channel} & \multicolumn{2}{c|}{1 lepton channel} & \multicolumn{2}{c}{2 leptons channel} \\
& \multicolumn{2}{c|}{} & $e$ sub-channel & $\mu$ sub-channel & $e$ sub-channel & $\mu$ sub-channel \\
\toprule
Trigger & \multicolumn{2}{c|}{ \met } &  Single electron & \met & Single electron & \met \\
\hline
Leptons & \multicolumn{2}{c|}{0 \VHloose\ leptons} & \multicolumn{2}{c|}{1 \WHsignal\ lepton} & \multicolumn{2}{c}{2 \VHloose\ leptons among which} \\
& \multicolumn{2}{c|}{ } &  $\pt>27$~\GeV & $\pt>25$~\GeV & \multicolumn{2}{c}{ $\geq$ 1 \WHsignal\ lepton, $\pt>27$~\GeV } \\
& \multicolumn{2}{c|}{ } &  \multicolumn{2}{c|}{no second \VHloose\ lepton}  & \multicolumn{2}{c}{both leptons of the same flavour} \\
& \multicolumn{2}{c|}{ } &  \multicolumn{2}{c|}{ }  & - & opposite sign muons  \\
 
\hline
\met & \multicolumn{2}{c|}{ $> 250$ \GeV} & $> 50$ \GeV & -  & \multicolumn{2}{c}{-} \\
\hline
\pTV & \multicolumn{6}{c}{$\pTV > 250$ \GeV } \\
\hline
\LargeR jets & \multicolumn{6}{c}{ at least one large-\textit{R} jet, $\pt> 250$ \GeV, $|\eta| < 2.0$} \\
\hline
Track-jets & \multicolumn{6}{c}{at least two track-jets, $\pt > 10$ \GeV,  $|\eta| < 2.5$, matched to the leading large-\textit{R} jet} \\
\hline
$b$-tagged jets & \multicolumn{6}{c}{leading two track-jets matched to the leading large-\textit{R} must be $b$-tagged (MV2c10, 70\%)} \\
\hline
\mJ      & \multicolumn{6}{c}{$> 50$ GeV} \\
\hline
$\min[\Delta \phi$(\vecmet, \smallR\ jets)] & \multicolumn{2}{c|}{ $> 30^{\circ}$} & \multicolumn{4}{c}{-} \\
\hline
$\Delta \phi$(\vecmet, $H_{\text{cand}}$) & \multicolumn{2}{c|}{ $> 120^{\circ}$} & \multicolumn{4}{c}{-} \\
\hline
$\Delta \phi$ (\vecmet, \vecmpt) & \multicolumn{2}{c|}{ $< 90^{\circ}$} & \multicolumn{4}{c}{-} \\
\hline
$\Delta y (V, H_{\text{cand}})$ & \multicolumn{2}{c|}{-} & \multicolumn{4}{c}{$|\Delta y (V, H_{\text{cand}})| < 1.4$} \\
\hline
\mll & \multicolumn{4}{c|}{-} & \multicolumn{2}{c}{ 66 \GeV $< \mll < 116$ \GeV} \\
\hline
Lepton \pT imbalance & \multicolumn{4}{c|}{-} & \multicolumn{2}{c}{$(\pT^{\ell_{1}} - \pT^{\ell_{2}} ) / \pTZ < 0.8$} \\
 
\bottomrule
\end{tabular}
\label{tab:EventSelection}
}
\end{table}

The multijet background in the 0-lepton channel originates mainly from
jet energy mismeasurements.  To reduce this background to a negligible
level, three dedicated selection criteria are applied.  Events are
removed if the missing transverse momentum is pointing towards the
direction of the Higgs-jet candidate ($\Delta \phi (\vecmet$,
$H_{\text{cand}})>120^{\circ}$). Events are also removed if the
calorimetric \vecmet\ and the track \vecmpt\ are far apart
($\Delta \phi (\vecmet, \vecmpt) < 90^{\circ}$).
The \vecmet\ is required to be isolated from any
calorimeter \smallR\ jet with transverse momentum in excess of
$70$~\GeV\ ($\min[\Delta \phi (\vecmet$, \smallR\ jets$)]>30^{\circ}$). In this case, only
\smallR\ jets not overlapping with the Higgs-jet candidate within $\Delta R=1.0$ are considered.

In the 1-lepton channel, the isolation requirements remove most of the
non-prompt lepton background.  An additional \met\ requirement is
applied in the electron sub-channel to reduce this background
further. In order to reduce other backgrounds, such as top and
$W+$jets production, a further selection on the rapidity difference
between the Higgs-jet candidate and the vector boson is applied
($|\Delta y (V,H_{\text{cand}})| < $ 1.4).
The $W$-boson rapidity is estimated assuming that \met\ is the \pt\ of
the neutrino and the longitudinal momentum of the neutrino is
estimated using the $W$-boson mass constraint. This method leads to a quadratic equation for the
longitudinal momentum of the neutrino. In case of two real solutions: the retained solution is the one that minimises the
difference between the longitudinal boost of the $W$ boson and the
Higgs boson. In case of no real solution, the imaginary part is set to 0.\footnote{This procedure is
equivalent to setting the reconstructed $W$ transverse mass to the $W$
mass.}

In the 2-lepton channel, where two same-flavour leptons are required
(in the dimuon sub-channel the two muons are further required to be of
opposite sign, in the dielectron case this selection is not applied
due to the comparatively higher charge misidentification), the
rapidity difference ($|\Delta y (V,H_{\text{cand}})| < $ 1.4)
effectively reduces the main $Z+$jets background.  A requirement is
imposed on the lepton \pt\ imbalance ($(\pT^{\ell_{1}}
- \pT^{\ell_{2}} ) / \pTZ < $ 0.8), which is sensitive to the
$Z$~boson polarisation~\cite{Goncalves:2018fvn}. Since the $Z$~boson
has different states of polarisation in the $ZH$ signal and the
$Z+$jets background, this selection further reduces this background.

Since the signal-to-background ratio increases for large Higgs boson
transverse
momenta~\autocite{Butterworth:2008iy,ATL-PHYS-PUB-2009-088}, events
are further split into two \ptv\ bins with $250<\ptv<400\,\GeV$ and
with $\ptv \geq 400\,\GeV$.
 
The selection efficiency in the 0-, 1- and 2- lepton channels and
two \ptv\ bins ranges between approximately 6\% and 16\% for the $WH$
and $ZH$ processes where the $W$ and $Z$ bosons decay leptonically and
the Higgs boson decays into a pair of $b$-quarks. The analysis does
not explicitly select $\tau$-leptons but they are accounted for in the
case of leptonically decaying $\tau$-leptons in the 1- and 2-lepton
channels and hadronically decaying $\tau$-leptons in the 0-lepton
channel if they are misidentified as jets.
 
As discussed in Section~\ref{sec:intro} the overlaps between the event
selections presented herein and those of Ref.~\cite{CompanionPaper}
are non negligible. In the 250~\GeV\ $<$ \ptv\ $<$ 400~\GeV\ region,
approximately 40\% of the signal events are selected by both sets of
selections, and the fraction of signal events uniquely selected by
the \largeR\ jet analysis varies between 5\% and 30\% with
increasing \ptv. In the \ptv\ $>$ 400~\GeV\ region, the overlap
decreases progressively to reach approximately 15\% and the
unique \largeR\ jet analysis signal events increase to 75\%\ at
a \ptv\ of around 700~\GeV.
 
The $t\bar{t}$ process is a major background in the 0- and 1-lepton
channels. For $t\bar{t}$ events, the $b$-tagged track-jets associated
with the Higgs-jet candidate are mainly a $b$- and a $c$-jet
(the former from a top-quark decay and the latter from the hadronic
$W$ boson decay) and therefore the second $b$-jet from the other top-quark
is often expected to be identified as an additional $b$-tagged track-jet not
associated with the Higgs-jet candidate. Taking this into account, signal regions (SR) in the 0- and
1-lepton channels are defined by vetoing on $b$-tagged track-jets
outside the Higgs-jet candidate and control regions (CR), enriched in
$t\bar{t}$ events, are built from events which fail this veto. The SRs
and CRs are accounted for in the same way in the fit, but CRs are
dominated by backgrounds and are used to constrain specific background
components.

Events in the 0- and 1-lepton channels are further categorised
depending on the number of \smallR\ jets not matched to the Higgs-jet
candidate, i.e. with $\Delta R$($H_{\text{cand}}$, \smallR\
jet)>1.0. Two categories are defined: a \textit{high-purity} signal
region (HP SR) with 0 \smallR\ jets not matched to the Higgs-jet
candidate and a \textit{low-purity} signal
region (LP SR) with $\geq 1$ \smallR\ jets not matched to the
Higgs-jet candidate.
 
The ten SRs and the four CRs are summarised in
Table~\ref{tab:SR_CR_definition}.

\begin{table}[!htbp]\small
\setlength{\tabcolsep}{2pt} \centering \caption{Summary of the
definition of the analysis regions. Signal enriched regions
are marked with the label SR. There are regions with
relatively large signal purity (HP SR) and with low purity (LP
SR). Background enriched regions are marked with the label
CR. The shorthand ``add'' stands for additional \smallR\ jets,
i.e. number of \smallR\ jets not matched to the Higgs-jet
candidate.}  \setlength{\extrarowheight}{1pt}
\label{tab:SR_CR_definition}
\begin{center}
 
\begin{tabular}{ l | c | c | c | c | c | c}
\toprule
\multirow{3}{*}{Channel}  & \multicolumn{6}{c}{Categories} \\
\cline{2-7}  & \multicolumn{3}{c|}{$250<\ptv<400$~\GeV} & \multicolumn{3}{c}{$\ptv \geq 400$~\GeV} \\
\cline{2-7}  & \multicolumn{2}{c|}{\shortstack{0 add. $b$-track-jets }} & \multirow{2}{*}{\shortstack{$\geq1$ add. \\$b$-track-jets}} &  \multicolumn{2}{c|}{\shortstack{0 add. $b$-track-jets}} &  \multirow{2}{*}{\shortstack{$\geq1$ add.\\ $b$-track-jets}}\\
 
\cline{2-3}
\cline{5-6}
& \shortstack{0 add. \\ small-\textit{R} jets}  & \shortstack{ $\geq1$ add.\\ small-\textit{R} jets}  &  &  \shortstack{0 add. \\ small-\textit{R} jets}  & \shortstack{$\geq1$ add. \\ small-\textit{R} jets} &  \\
\toprule
0-lepton      & HP SR  & LP SR & CR   & HP SR & LP SR  & CR   \\
\hline
1-lepton      & HP SR  & LP SR & CR   & HP SR & LP SR  & CR   \\
\hline
2-lepton & \multicolumn{3}{c|}{SR} & \multicolumn{3}{c}{SR} \\
\bottomrule
\end{tabular}
\end{center}
 
\end{table}

 
\section{Background composition and estimation}
\label{sec:background}
The background contribution in the SRs is different for each of the
three channels. In the 0-lepton channel, the dominant
background sources are \zjets\ and \tt\ events with a significant
contribution from \wjets\ and diboson production. In the 1-lepton
channel, the largest backgrounds are \tt\ and \wjets\ production
followed by the single-top background. In the 2-lepton
channel, \zjets\ production is the dominant background followed by the
$ZZ$ background. Contributions from $t\bar{t}V$ and $t\bar{t}H$
are negligible. The multijet background, due to semileptonic
heavy-flavour-hadron decays or misidentified jets, is found to be
negligible in the 0- and 2-lepton channels as well as in the 1-lepton
muon sub-channel after applying the event selections described in
Section~\ref{sec:selection}, as confirmed using data-driven
techniques. In the 1-lepton electron sub-channel its contribution is
not neglected. All initial background distribution shapes prior to the
fit (described in Section~\ref{sec:results}), except those for
multijet, are estimated from the samples of simulated events. The
multijet shape and normalisation are determined using data.
 
The $W/Z+$jets simulated event samples are split into 6 categories
depending on the `truth' labels of the track-jets ghost-associated to
the Higgs-jet candidate: $W/Z+bb$, $W/Z+bc$, $W/Z+bl$, $W/Z+cc$,
$W/Z+cl$ and $W/Z+ll$; in this notation $l$ refers to a light-flavour
jet.\footnote{When labelling jets in the \vjets\ backgrounds
modelling, the labelling of $\tau$-jets is omitted and the negligible
$\tau$-lepton contribution is included with light-flavour jets.} The
$W/Z+bb$ fraction corresponds to approximately 80\% of the total
$W/Z+$jets background. This categorisation is used in the
uncertainties variations of the ratios $V+bc/V+bb$, $V+bl/V+bb$ and
$V+cc/V+bb$ to cover uncertainties on the flavour composition in
$V$+jets production, see Section~\ref{sec:systematics}.
 
In the statistical analysis
described in Section~\ref{sec:results}, the components $W/Z+bb$,
$W/Z+bc$, $W/Z+bl$ and $W/Z+cc$ are treated as a single background
component denoted by $W/Z$+HF. The $W$+HF and $Z$+HF contributions,
which together constitute 90\% of $V+$jets background, are estimated
separately, each with its own normalisation factor determined from the
fit to data.

The $t\overline{t}$ production background arises from topologies with
decays of $W$ bosons into $\tau$-leptons which then decay hadronically in
the 0-lepton channel and from $W$ bosons decaying into $e/\mu$ in the
1-lepton channel. In the 2-lepton channel the $t\overline{t}$
contribution is much smaller. For the 0- and 1-lepton channels, two
independent normalisation factors are considered and left floating in
the fit, where they are constrained by the CRs.
 
Single-top production contributes to the 0- and 1-lepton channels
and $Wt$ production is the dominant process ($s$- and $t$-channel
processes amount to less than 1\% globally and less than 5\% of the single-top contribution).
 
The diboson background process consists of final states arising mostly from $WZ$
and $ZZ$ events, where a $Z$ boson decays into a pair of
$b$-quarks. This process has a topology very similar to that of the
signal, exhibiting a peak in $m_\textrm{J}$ at the mass of the
hadronically decaying vector boson. Although it is a subdominant
contribution, it provides an important reference for validation. Its
normalisation is measured simultaneously with the $VH$ signal.

In the 1-lepton channel, the multijet background originating from jets
misidentified as leptons and/or due to semileptonic
heavy-flavour-hadron decays cannot be neglected. Since MC simulation
samples are statistically limited and are not expected to reproduce
the multijet production in this corner of the phase space, it is
estimated from a template fit using the data. The $m_\textrm{J}$
templates in the electron and muon sub-channels are taken from dedicated CRs
enriched in multijet background, obtained from the inversion of the
tight lepton isolation requirements and the removal of the \met\
requirement, and after subtraction of the other backgrounds.
The multijet normalisations are estimated in the SRs from a fit to the
transverse mass~\footnote{The transverse mass $m_\textrm{T}$ of the
$W$ boson candidate in the event is calculated using the lepton
candidate and
\met\ according to $m_\textrm{T} = \sqrt{2 p_\textrm{T}^{\ell} \met (1-\cos{\Delta \phi(\ell,\met))}}$.} distribution separately for the electron and muon
sub-channels. The contribution of the multijet background is found to
be negligible in the muon sub-channel. In the electron sub-channel it
is approximately 2\%\ of the total background, with an uncertainty of
55\% estimated mainly from the statistical uncertainty of the transverse mass fit.
This contribution and its associated uncertainty are taken into
account in the signal extraction fit.

 
\section{Systematic uncertainties}
\label{sec:systematics}
Systematic uncertainties can have an impact on the
overall signal and background yields, on the shapes of the jet mass
distributions, on the CR to SR extrapolations, and on the relative
acceptances between the HP and LP SRs and between the \pTV\ bins.
Systematic uncertainties are discussed herein for three main
categories: experimental, signal modelling, and background
modelling.

\subsection{Experimental systematic uncertainties}
 
The uncertainties in the \smallR\ jet energy scale and resolution have
contributions from $in$ $situ$ calibration studies, from the dependency on
the pile-up activity and on the flavour composition of
the jets~\cite{PERF-2016-04}. For \largeR\ jets, the uncertainties in the
energy and mass scales are based
on a comparison of the ratio of calorimeter-based to track-based
measurements in dijet data and simulation, as described in
Ref.~\cite{Aaboud:2018kfi}. The impact of the jet energy scale and
resolution uncertainties on the \largeR\ jet mass are assessed by
applying different calibration scales and smearings to the jet
observables in the simulation, according to the estimated
uncertainties. An absolute uncertainty of 2\% is used for the jet
energy resolution while a relative uncertainty of 20\% is used for the
jet mass resolution, consistent with previous studies for trimmed
jets~\cite{STDM-2011-19, ATLAS-CONF-2016-035}.

The $b$-tagging uncertainties are assessed from the calibration data
in various kinematic regions and separately for $b$-, $c$-, and
light-flavour jets. The uncertainties are then decomposed in each of
the flavour categories into independent components. An additional
uncertainty is included to account for the extrapolation to jets
with \pt\ beyond the kinematic reach of the data calibration (the
thresholds are 250~\GeV, 140~\GeV\ and 300~\GeV\ for $b$-, $c$- and
light-flavour jets,
respectively)~\cite{PERF-2016-05,ATLAS-CONF-2018-006,ATLAS-CONF-2018-001}.
 
Other experimental systematic uncertainties with a smaller impact are those
in the lepton energy and momentum scales, in lepton reconstruction and
identification efficiency, and in the efficiency of the triggers.
An uncertainty associated with the modelling of \pileup in the
simulation is included to cover the difference between the predicted
and measured inelastic cross-sections~\cite{STDM-2015-05}. The
uncertainties in the energy scale and resolution of the \smallR\ jets
and leptons are propagated to the calculation of \met, which also has
additional uncertainties from the scale, resolution and reconstruction
efficiency of the tracks used to compute the soft term, along with the
modelling of the underlying event~\autocite{PERF-2016-07}.
 

\subsection{Signal modelling systematic uncertainties}

The systematic uncertainties that affect the modelling of the signal
are derived closely following the procedure outlined in
Refs.~\autocite{HIGG-2016-29,HIGG-2018-50, deFlorian:2016spz} and in
Refs.~\cite{ATL-PHYS-PUB-2018-035, Bendavid:2018nar} for uncertainties specific to
STXS. The systematic uncertainties in the calculations of the $VH$
production cross-sections and the $H\to\bbbar$ branching fraction are
assigned following the recommendations of the LHC Higgs Cross Section
Working
Group~\cite{Dittmaier:2011ti,Dittmaier:2012vm,Heinemeyer:2013tqa,Harlander:2013mla,Brein:2012ne}.
Acceptance and shape systematic uncertainties are derived to account
for missing higher-order QCD and EW corrections, for
PDF+$\alpha_{\textrm S}$ uncertainties, and for variations of the PS
and UE models. Factorisation and renormalisation scales
are varied by factors of 0.5 and 2. PDF-related uncertainties are derived following Ref.~\cite{Butterworth:2015oua}. The effects of the uncertainties from missing
higher-order EW corrections, PDF+$\alpha_{\textrm S}$ and QCD scale
variations on the jet mass shape are negligible. The PS and
UE uncertainty is evaluated by comparing the nominal
signal \POWHEGBOX  samples showered by \textsc{Pythia 8} with
alternative samples showered
by \textsc{Herwig~7}~\cite{Bellm:2015jjp}.

\subsection{Background modelling systematic uncertainties}

The principal additional modelling uncertainties for the backgrounds
that were considered are the following: renormalisation and
factorisation scale variations by factors of 0.5 and 2 for higher
order in QCD corrections of the matrix element of the process; merging
scale variations from multi-leg simulations; resummation scale or
parton shower uncertainties; PDF uncertainties; and differences with
alternative MC generators. The impact of these systematic
uncertainties in terms of normalisation, shape, acceptance and
extrapolation between analysis regions is then estimated and included
in the fit model (described in Section~\ref{sec:results}). Given that
the analysis is based on the fit of the $m_\textrm{J}$ variable only,
all shape uncertainties are estimated with respect to this observable.
 
The normalisations of the $W/Z$+HF backgrounds are free parameters in the
fit. They are determined thanks to the use of the jet mass distributions in SRs
once $t\bar{t}$ is constrained from $CR$ enriched in $t\bar{t}$ events. In addition to scale variations within
\textsc{Sherpa 2.2.1}, alternative samples for acceptance and shape
variations generated with \textsc{MadGraph} interfaced to \textsc{Pythia 8} were
considered. Finally, variations in the
$V+bc/V+bb$, $V+bl/V+bb$ and $V+cc/V+bb$ ratios are
accounted for independently for the $W$- and $Z$-boson backgrounds.

For top-quark pair production modelling uncertainties, specific
initial-state radiation (ISR) and final-state radiation (FSR)
\textsc{Pythia} parameters are used to assess the related systematic
uncertainties. In addition to the typical scale variations,
alternative NLO samples using the \MGMCatNLO
and
\textsc{Herwig 7} generators were considered.  The
$t\overline{t}$ normalisation is free in the fit and mainly
constrained in the CRs for the 0- and 1-lepton channel. For the
2-lepton channel it is constrained to its nominal predicted value
with an uncertainty of 20\%. Due to top decays not fully contained within the \largeR\ jet,
the relative number of events where exactly two and where three or
more VR track-jets are ghost-associated to the \largeR\ jet can modify
the \largeR\ jet mass template. This is accounted for by an additional uncertainty
estimated from the impact on the $t\overline{t}$ background template of a 20\% variation in this relative ratio.

The normalisations, acceptances and shapes of all single-top
production processes are constrained to their predictions within
the corresponding uncertainties.
For the dominant $Wt$ channel, ISR/FSR uncertainties as well as
alternative generator samples, \textsc{Herwig 7} and
\textsc{Madgraph5\_aMC@NLO}, are considered. Since the $Wt$ channel
has the same flavour composition and a similar shape in the 0- and
1-lepton channels, the modelling uncertainties were studied in the
1-lepton channel and then propagated to the 0-lepton channel. An
associated extrapolation uncertainty is taken into account.
 
To account for the ambiguities in the interference between \ttbar\ and
single-top production, an alternative sample generated with \POWHEGBOX
interfaced to \textsc{Pythia}~8, using the diagram subtraction (DS)
scheme, is used~\cite{Frixione:2008yi}.  The difference between the DS
and DR schemes for the $Wt$ single-top production is accounted for as
an additional systematic uncertainty.

For diboson production, in addition to the scale variations for
acceptance, extrapolation and shape systematic uncertainties,
alternative diboson samples were generated using \POWHEGBOX
interfaced to \textsc{Pythia 8} and the difference with respect to the \textsc{Sherpa} nominal samples was used as an additional uncertainty.


\section{Results}
\label{sec:results}
 
The results are obtained from a binned maximum-profile-likelihood fit
to the data of the \mJ distribution, using all the signal and control
regions defined in Section~\ref{sec:selection}.  The fit is performed
using the \textsc{RooStats} framework~\cite{RooFit,
2010acatconfE}. Signal and background \mJ templates are determined
from MC simulation (described in Section~\ref{sec:dataMC}) in all
cases except for the multijet background in the 1-lepton channel,
which is extracted from the data as discussed in
Section~\ref{sec:background}.
 
\begin{figure}[htbp]
\centering
\subfloat[]{\includegraphics[width=0.35\textwidth]{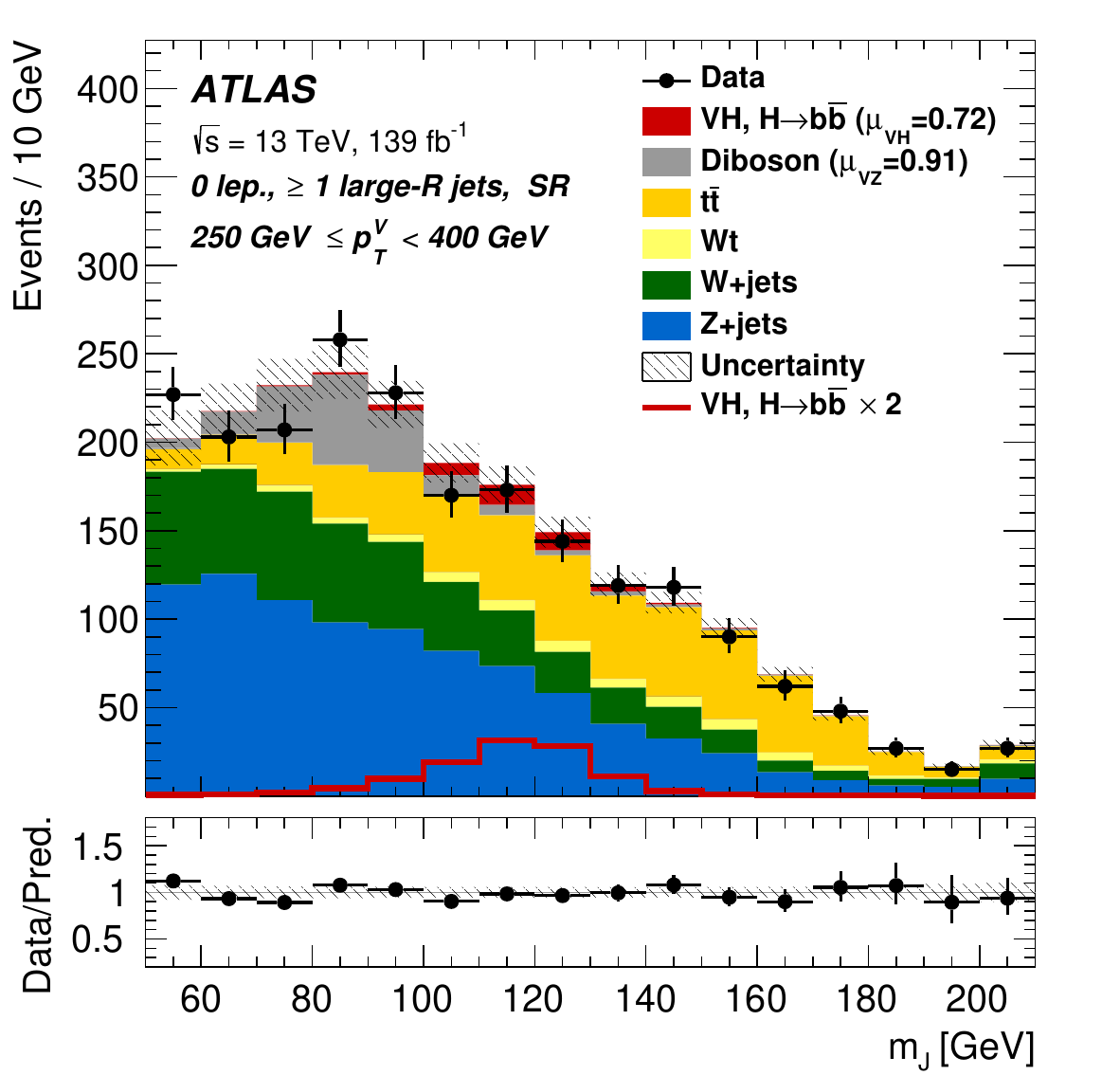}}
\subfloat[]{\includegraphics[width=0.35\textwidth]{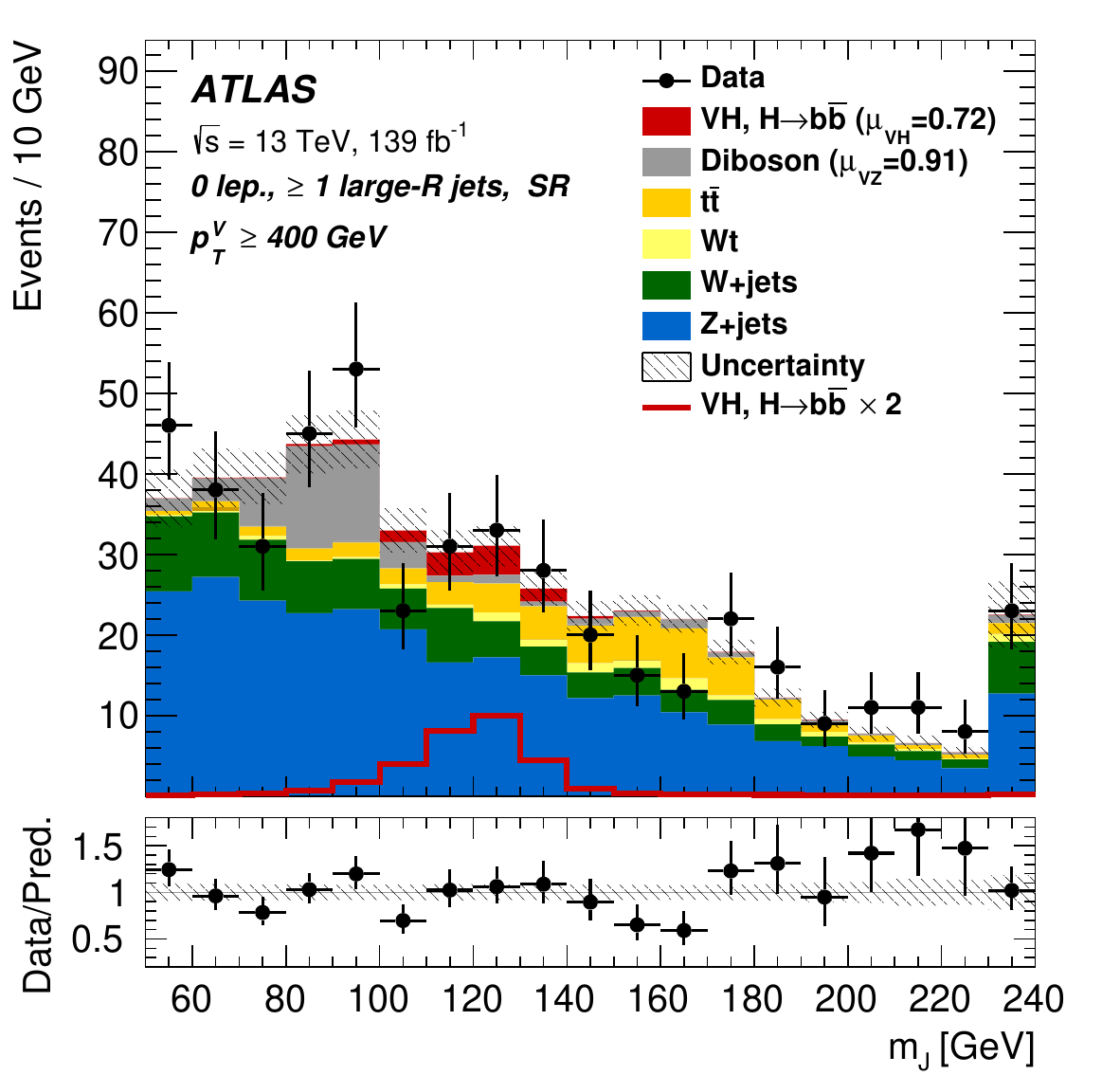}}
\\
\subfloat[]{\includegraphics[width=0.35\textwidth]{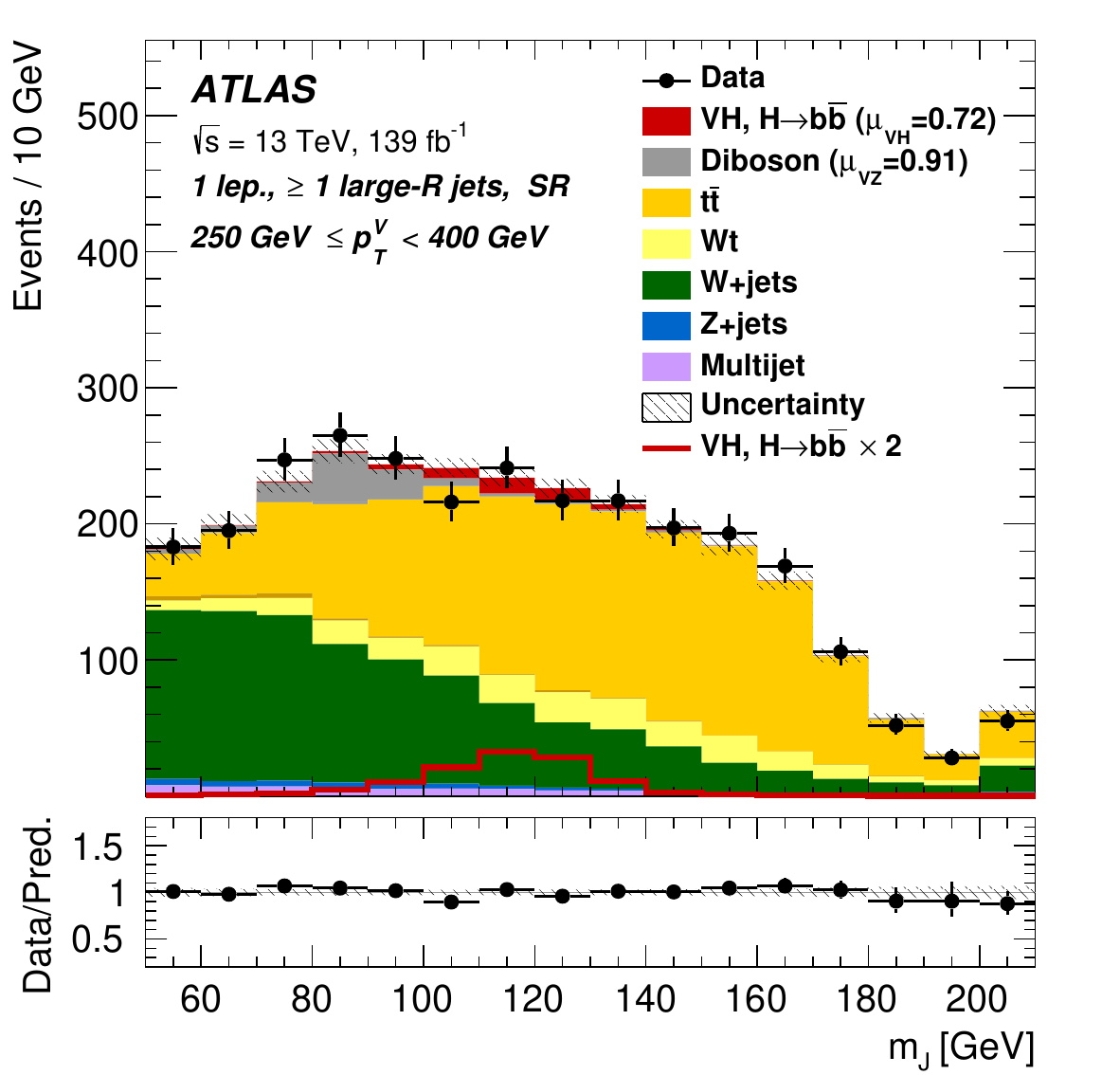}}
\subfloat[]{\includegraphics[width=0.35\textwidth]{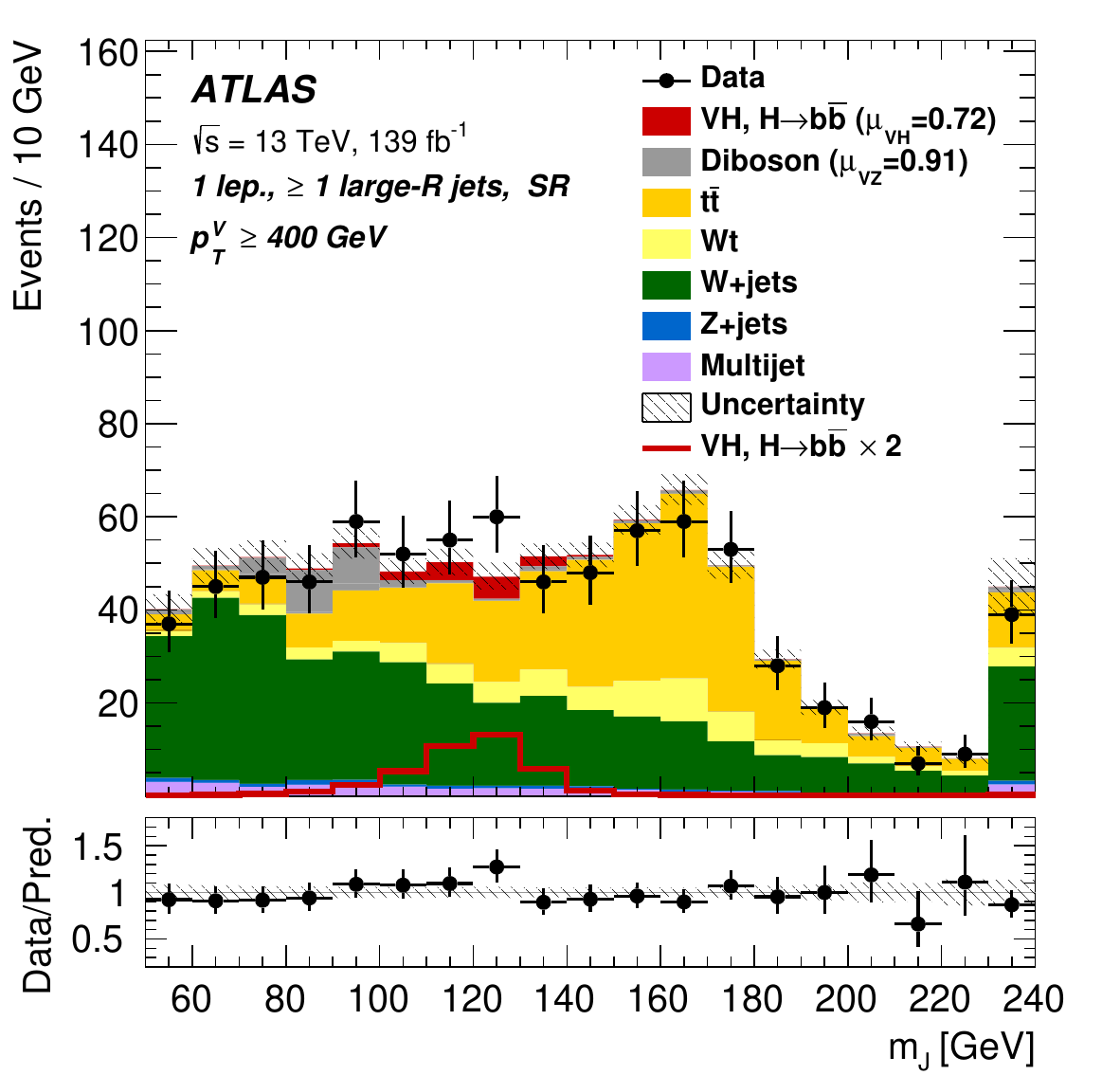}}
\\
\subfloat[]{\includegraphics[width=0.35\textwidth]{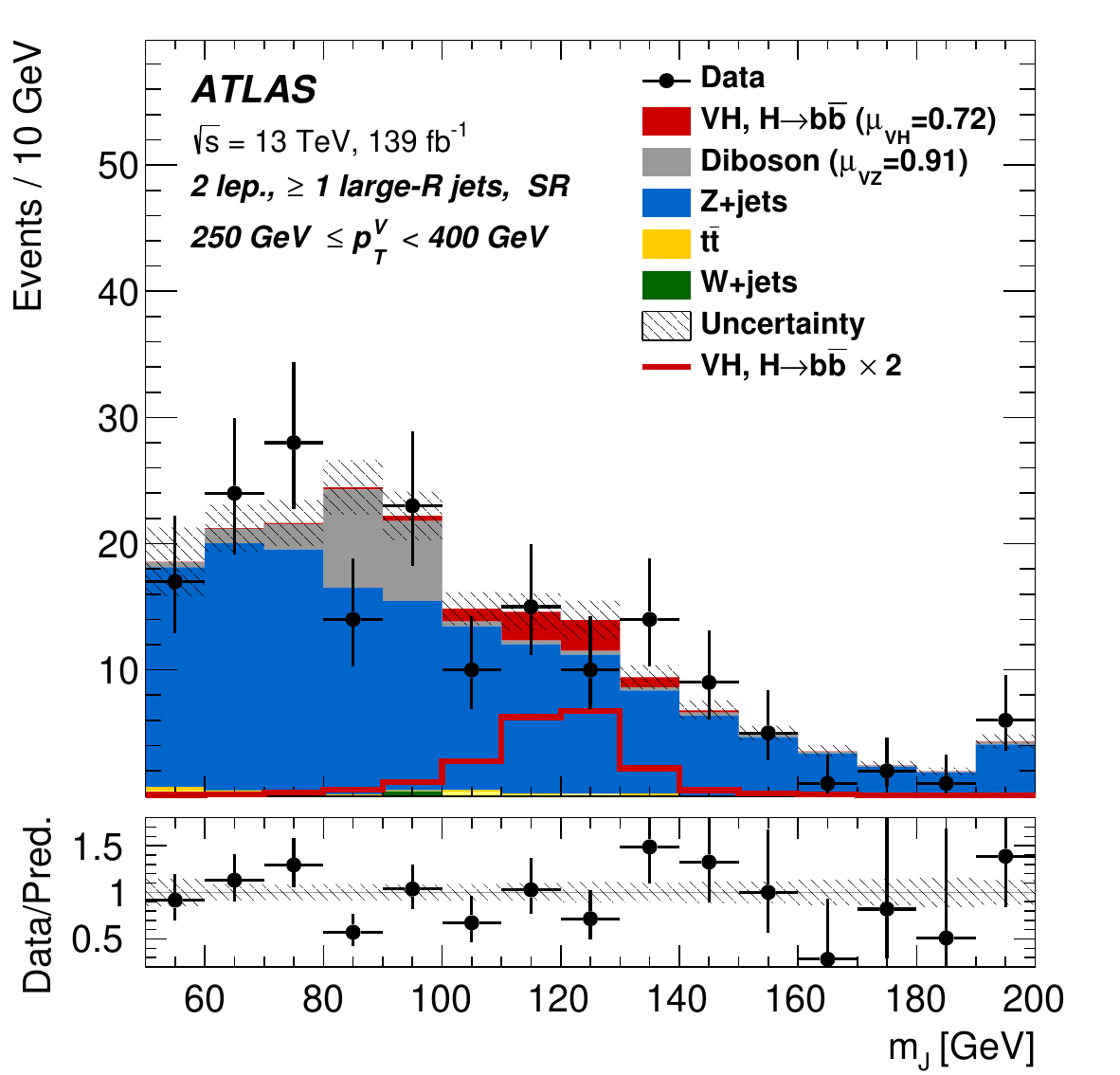}}
\subfloat[]{\includegraphics[width=0.35\textwidth]{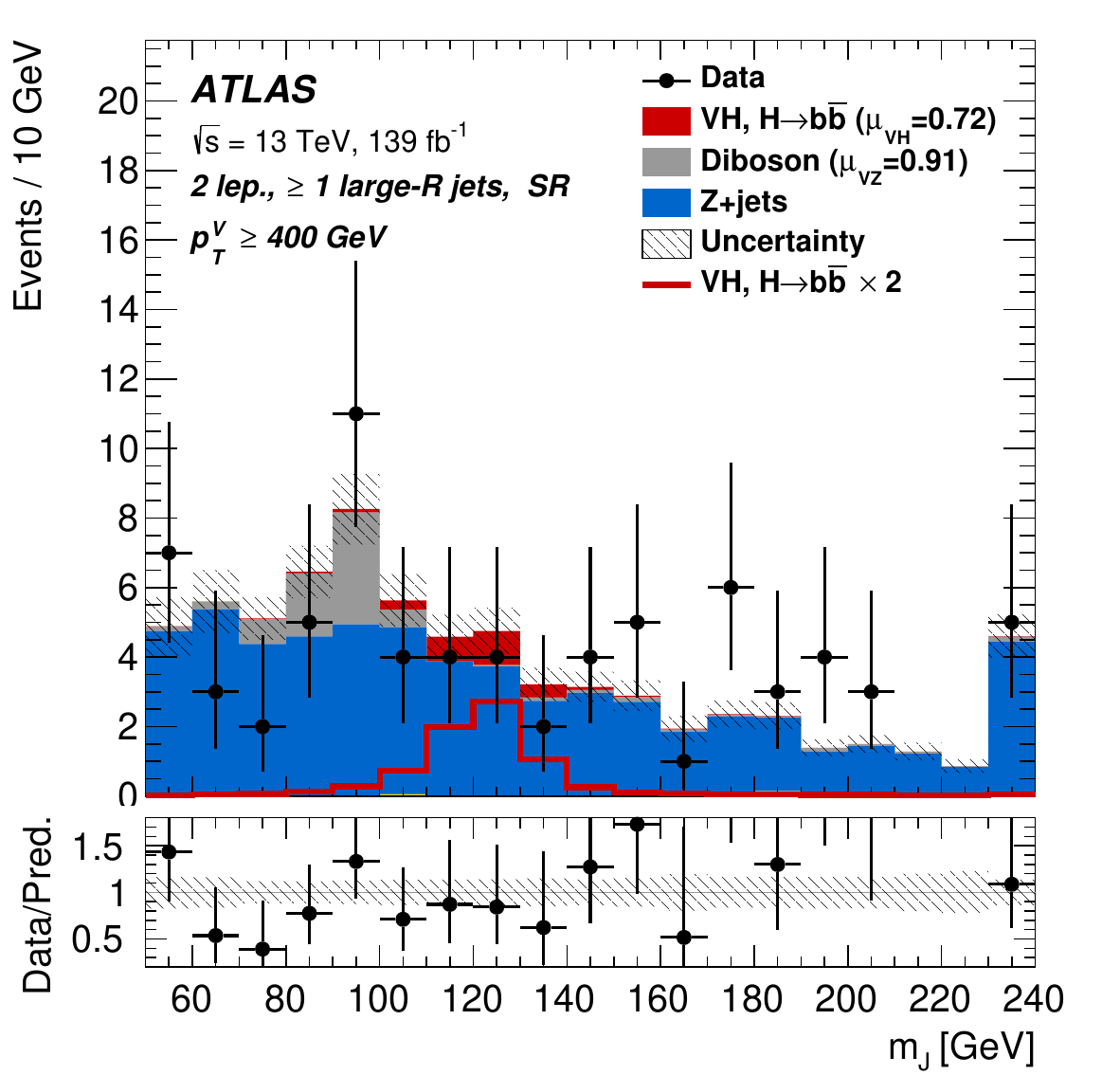}}
\\
\caption{The \mJ\ post-fit distributions
in (a, b) the \zlep, (c, d) \olep and (e, f) \tlep signal regions for
2-$b$-tagged events for (a, c, e) $250\,\GeV<\ptv<400\,\GeV$ and (b,
d, f) $\ptv \geq 400\,\GeV$. The low-purity and high-purity
categories in the case of the \zlep and \olep channels are merged in
this figure. The background contributions after the likelihood fit
are shown as filled histograms. The Higgs boson signal ($\mh =
125$~\GeV) is shown as a filled histogram on top of the fitted
backgrounds normalised to the signal yield extracted from data
($\mu_{VH}^{bb}=0.72$), and unstacked as an unfilled histogram,
scaled by the SM prediction times a factor of two. The size of the
combined statistical and systematic uncertainty for the sum of the
fitted signal and background is indicated by the hatched band. The
highest bin in the distributions contains the overflow. The ratio of
the data to the sum of the fitted signal and background is shown in
the lower panel.
\label{fig:jetmass_postfitoutput}}
\end{figure}
 
The likelihood function is constructed from the product of the Poisson
probabilities of each bin of the mass distributions and auxiliary
terms used to model systematic uncertainties. The likelihood function
is described in more detail in Ref.~\autocite{HIGG-2016-29}.  The
parameters of interest (POI) are the signal strengths $\mu$,
multiplication factors that scale the expected SM Higgs boson signal,
in one or more sub-channels, or the $VZ$ process. The signal strength
parameters are extracted simultaneously with the overall diboson signal
strength $\mu_{VZ}^{bb}$ by maximising the likelihood.

Systematic uncertainties are modelled in the likelihood function by
parameterised variations of the number of signal and background events,
and of the templates through nuisance parameters (NP). Systematic
variations of the templates that are subject to large statistical
fluctuations are smoothed, and systematic uncertainties that have a
negligible impact on the final results are pruned away
region-by-region~\cite{HIGG-2013-23}.  NPs corresponding to most
uncertainties discussed in Section~\ref{sec:systematics} are
constrained using Gaussian or log-normal probability density functions
as auxiliary terms in the likelihood. The normalisations of the
largest backgrounds, $t\bar{t}$ (in the 0- and 1-lepton channels),
$W$+HF and $Z$+HF, are left unconstrained in the fit. The background normalisation factor values from the fit correspond to $0.88\pm0.10$ and $0.83\pm0.09$ for $t\bar{t}$, in the 0- and 1-lepton channels, respectively; $1.12\pm0.14$ for $W$+HF and $1.32\pm0.16$ for $Z$+HF and are also summarised in Table~\ref{tab:resultscalefactors}. The fit model uses a single normalisation factor for $Z$+HF and compatible results were found when using two different factors for the 0- and 2-lepton channels. To account for the uncertainty due to the limited size of the MC simulation samples, an NP is used for each bin of the templates~\cite{BARLOW1993219}.
 
\begin{table}[hptb]
\begin{center}
\caption{Factors applied to
the nominal normalisations of the \ttb, \whf\, and \zhf\ backgrounds, as obtained from the likelihood fit. The errors represent the combined statistical and systematic uncertainties.
\label{tab:resultscalefactors}}
\begin{tabular}{lc}
\toprule
Process and Category & Normalisation factor \\
\midrule
\ttb\ 0-lepton & $0.88 \pm 0.10$ \\
\ttb\ 1-lepton & $0.83 \pm 0.09$ \\
\whf\  &  $1.12 \pm 0.14$ \\
\zhf\ &  $1.32 \pm 0.16$ \\
\bottomrule
\end{tabular}
\end{center}
\end{table}

The \mJ distributions with signal strengths, background normalisations
and all NPs set at their best-fit values, are shown in
Figure~\ref{fig:jetmass_postfitoutput} for all three channels' SRs and
in Figure~\ref{fig:jetmass_postfitoutput_CR} for the CRs.
The low-purity and high-purity categories in the case of the \zlep and \olep channels are merged in Figure~\ref{fig:jetmass_postfitoutput}.  In all SRs and CRs a good agreement between the data and the prediction is observed.

\begin{figure}[ht]
\centering
\subfloat[]{\includegraphics[width=0.33\textwidth]{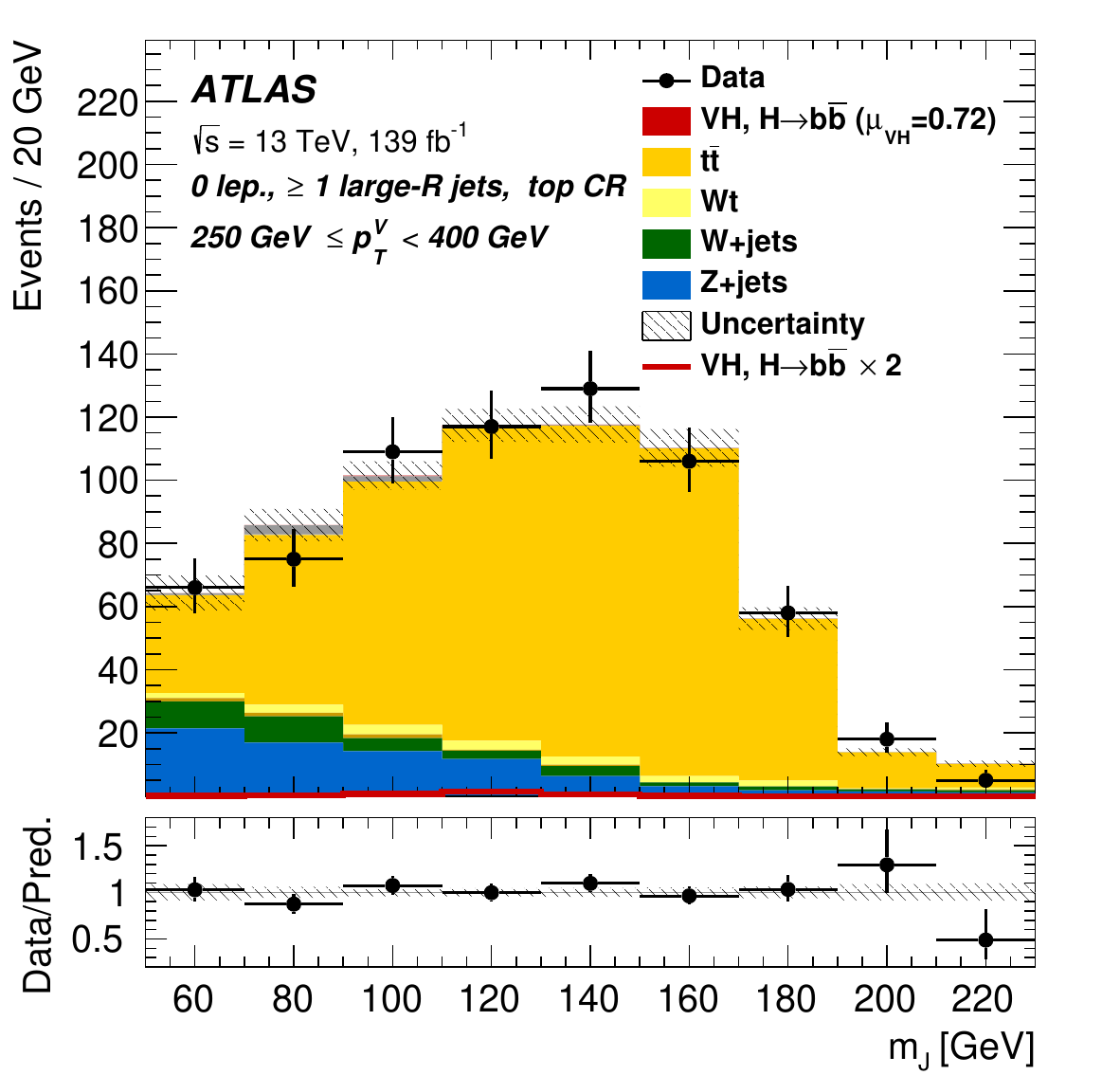}}
\subfloat[]{\includegraphics[width=0.33\textwidth]{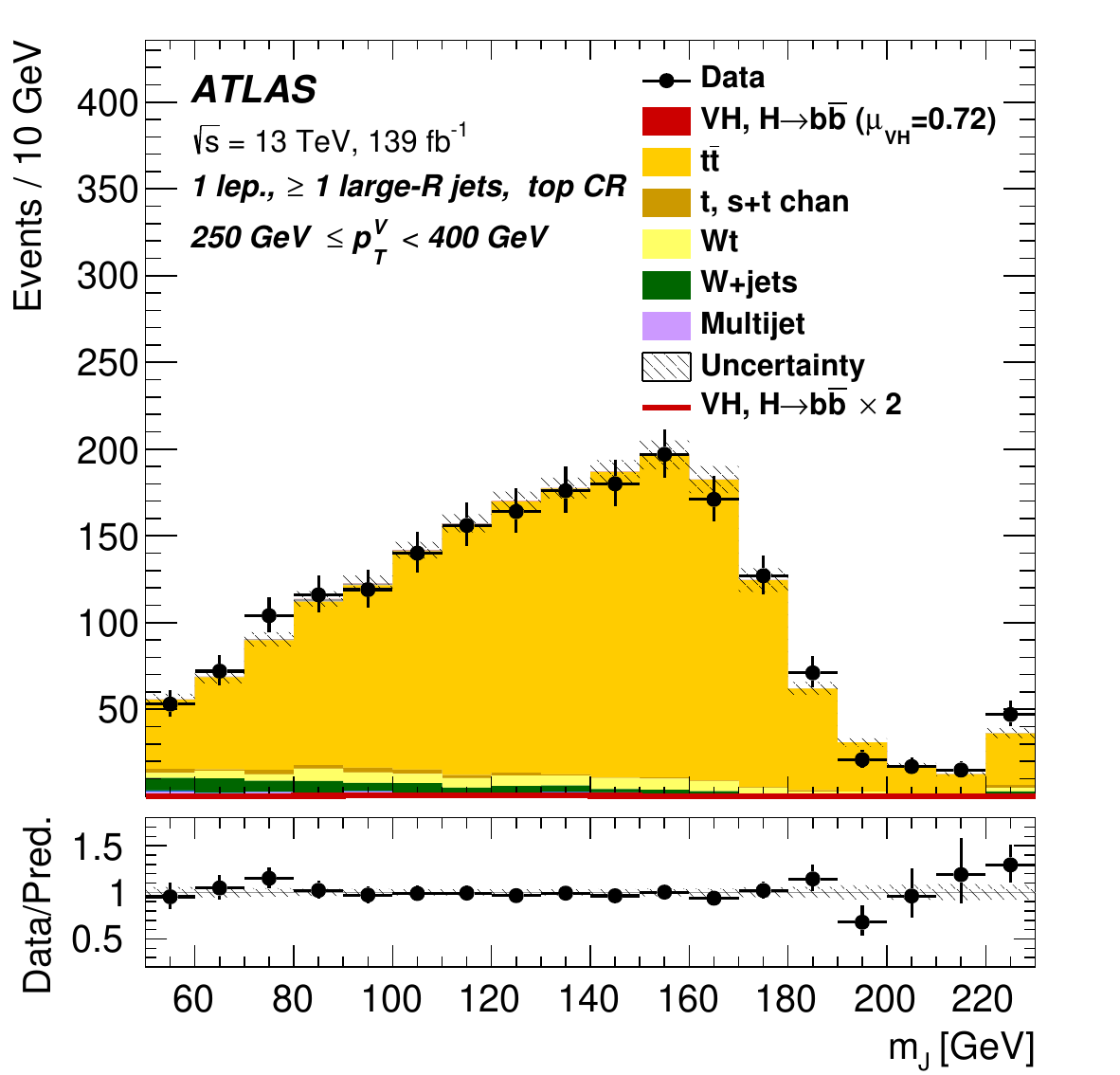}}\\
 
\subfloat[]{\includegraphics[width=0.33\textwidth]{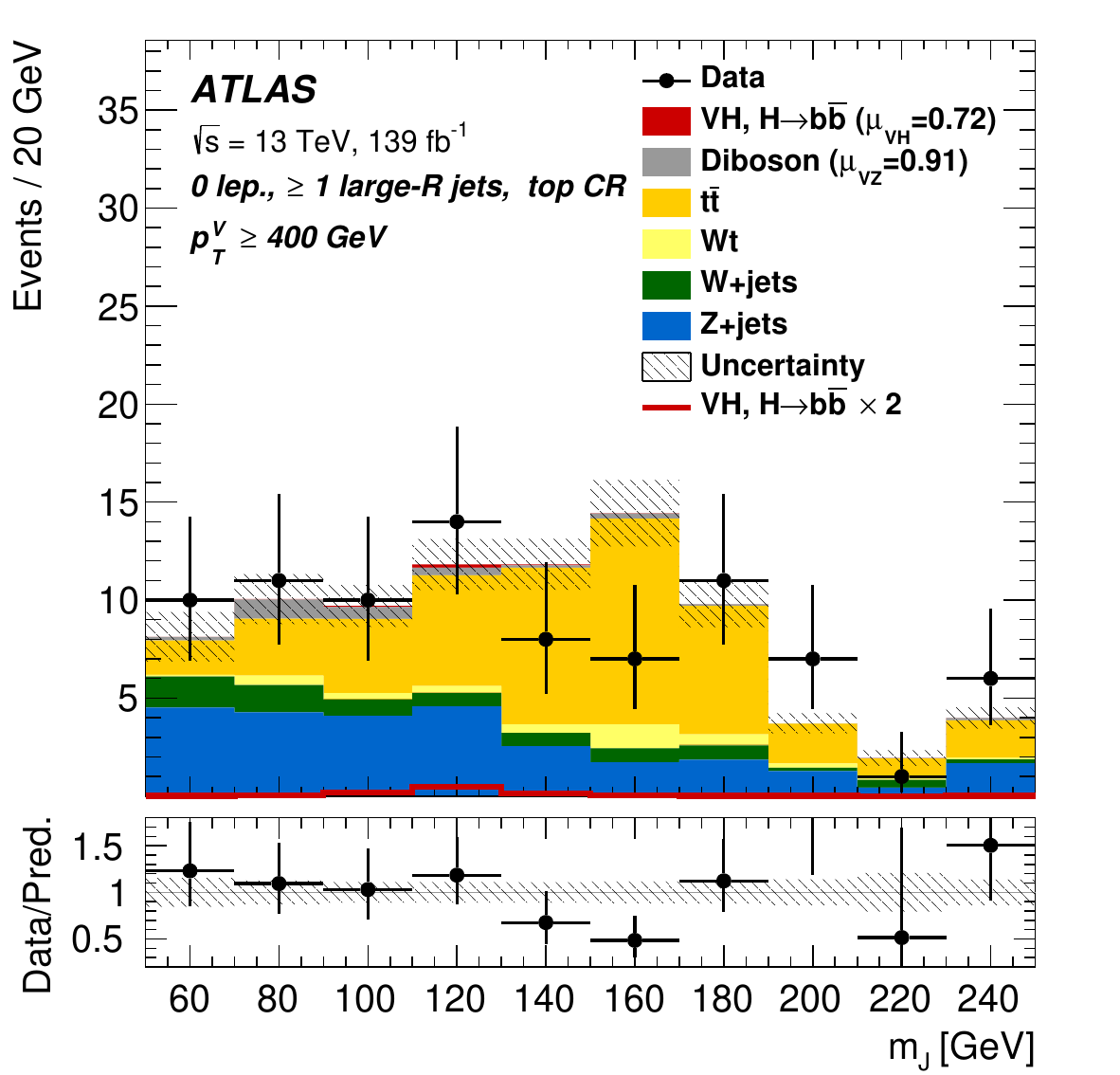}}
\subfloat[]{\includegraphics[width=0.33\textwidth]{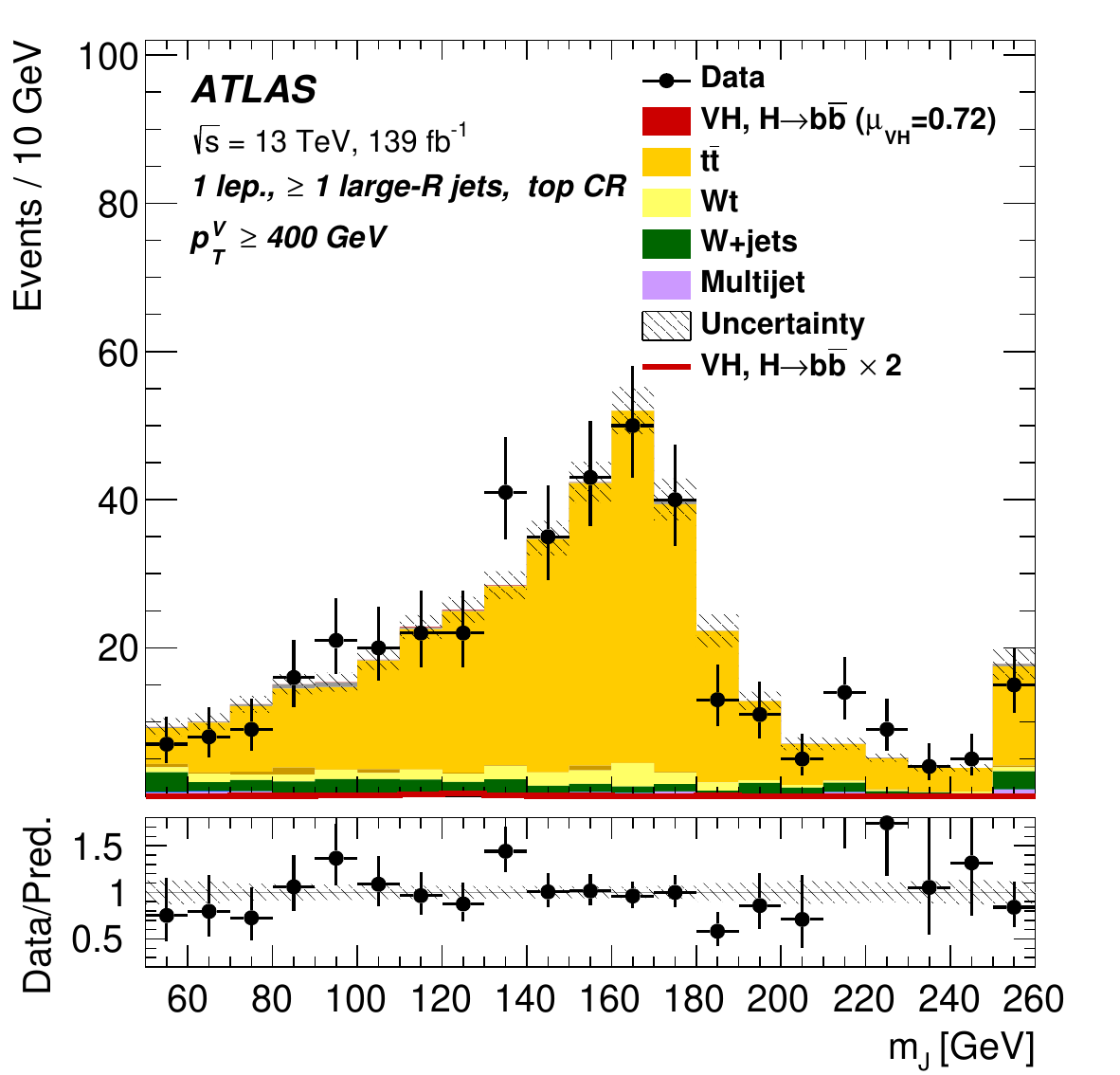}}

\caption{The \mJ\  post-fit distributions in the \ttbar\ control region for
(a, b) the \zlep\ channel and the \olep\ channel for $250\,\GeV<\ptv<400\,\GeV$ and (c, d)
the \zlep\ channel and the \olep\ channel for $\ptv>400\,\GeV$.  The background
contributions after the likelihood fit are shown as filled
histograms. The Higgs boson signal ($\mh = 125$~\GeV) is shown as a
filled histogram on top of the fitted backgrounds normalised to the
signal yield extracted from data ($\mu_{VH}^{bb}=0.72$), and
unstacked as an unfilled histogram, scaled by the SM prediction times a factor
of 2. The size of the combined statistical and systematic
uncertainty for the sum of the fitted signal and background is
indicated by the hatched band. The highest bin in the distributions
contains the overflow. The ratio of the data to the sum of the
fitted signal and background is shown in the lower panel.
\label{fig:jetmass_postfitoutput_CR}}
\end{figure}
 
For a Higgs boson mass of 125~\GeV, when all lepton channels are
combined, the observed excess with respect to the background-only
hypothesis has a significance of 2.1 standard deviations, to be compared with an
expectation of 2.7 standard deviations. The fitted $\mu_{VH}^{bb}$ value is:
\begin{eqnarray*}
\mu_{VH}^{bb} = 0.72 ^{+0.39}_{-0.36} = 0.72 ^{+0.29}_{-0.28}  \mathrm{(stat.)} ^{+0.26}_{-0.22} \mathrm{(syst.)}.
\end{eqnarray*}
 
In this result, the largest uncertainties are statistical and include
the impact from the floating background normalisations unconstrained
in the fit. The latter component is subdominant. The impact of
systematic uncertainties is almost as important as the total statistical uncertainty. The dominant source
of systematic uncertainty is experimental and related to the \largeR\
jet calibration, in particular in the \mJ\ resolution, amounting to
an impact of approximately 0.13 on $\mu_{VH}^{bb}$. The second largest
source of systematic uncertainty is the background modelling, which
overall has an impact of approximately 0.10 on the result. The limited
size of the MC simulation samples has a non-negligible impact of
0.09. Systematic uncertainties in the signal modelling have an impact
of approximately 0.04, on par with uncertainties related to \smallR\
jets. The breakdown of the systematic uncertainties in the measurement of the
signal strength are displayed in Table~\ref{tab:mu_syst_unc}.

\begin{table}[tb]
\caption{Breakdown of the absolute contributions to the uncertainty in $\mu_{VH}^{bb}$ inclusive in \pTV.
The sum in quadrature of the systematic uncertainties attached to the
categories differs from the total systematic uncertainty due to correlations. The reported values represent the average between the positive and negative uncertainties on $\mu_{VH}^{bb}$.
\label{tab:mu_syst_unc}}
\begin{center}
 
\begin{tabular}{ l | l  c c }
\toprule
\multicolumn{2}{l}{Source of uncertainty}  & Avg. impact \\
\toprule
\multicolumn{2}{l}{Total} & 0.372 \\
\multicolumn{2}{l}{Statistical} & 0.283 \\
\multicolumn{2}{l}{Systematic} & 0.240 \\
\midrule
\multicolumn{3}{l}{Experimental uncertainties}\\
\midrule
\multicolumn{2}{l}{\SmallR\  jets} &  0.038 \\
\multicolumn{2}{l}{\LargeR\  jets} &  0.133 \\
\multicolumn{2}{l}{\met} & 0.007 \\
\multicolumn{2}{l}{Leptons} & 0.010 \\
\multirow{3}{*}{$b$-tagging~~~} & $b$-jets & 0.016 \\
& $c$-jets &  0.011 \\
& light-flavour jets & 0.008 \\
& extrapolation & 0.004 \\
\multicolumn{2}{l}{Pile-up} & 0.001 \\
\multicolumn{2}{l}{Luminosity} & 0.013 \\
\midrule
\multicolumn{3}{l}{Theoretical and modelling uncertainties}\\
\midrule
\multicolumn{2}{l}{Signal} & 0.038 \\
\multicolumn{2}{l}{Backgrounds} & 0.100 \\
\multicolumn{2}{l}{$\hookrightarrow$ $Z$~+~jets} & 0.048 \\
\multicolumn{2}{l}{$\hookrightarrow$ $W$~+~jets} & 0.058 \\
\multicolumn{2}{l}{$\hookrightarrow$ $t\bar{t}$} & 0.035 \\
\multicolumn{2}{l}{$\hookrightarrow$ Single top quark} & 0.027 \\
\multicolumn{2}{l}{$\hookrightarrow$ Diboson} & 0.032 \\
\multicolumn{2}{l}{$\hookrightarrow$ Multijet} & 0.009 \\
\midrule
\multicolumn{2}{l}{MC statistical} & 0.092 \\
\bottomrule
\end{tabular}
\end{center}
\end{table}

The \mJ distribution is shown in
Figure~\ref{fig:jetmass_backgroundsubstracted} summed over all
channels and signal regions, weighted by their respective values of
the ratio of the fitted Higgs boson signal to background yields and after
subtraction of all backgrounds except for the $WZ$ and $ZZ$ diboson
processes.
 
\begin{figure}[ht!]
\begin{center}
\subfloat[]{\includegraphics[width=0.414\textwidth]{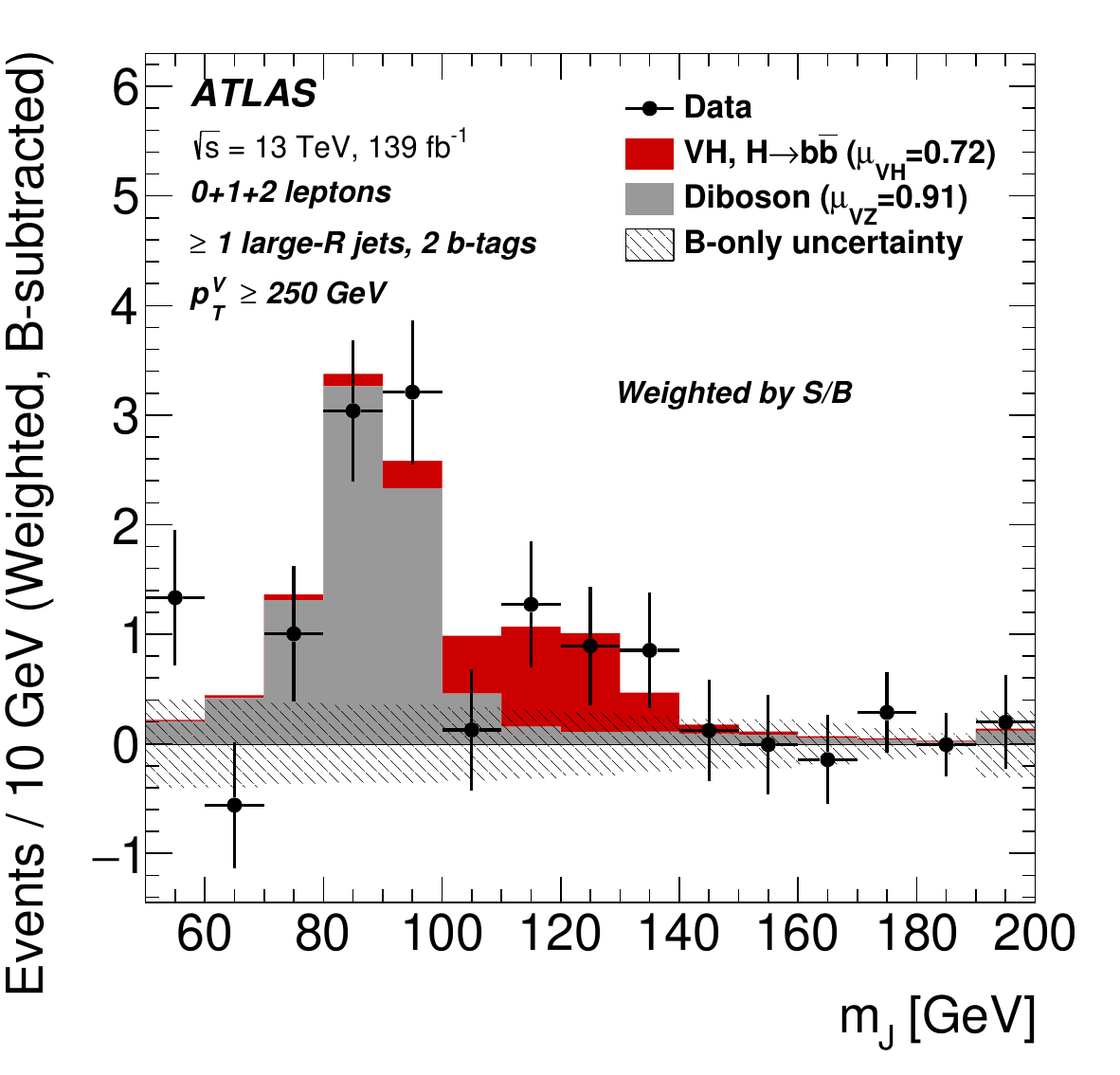}\label{fig:jetmass_backgroundsubstracted}}
\subfloat[]{\includegraphics[width=0.554\textwidth]{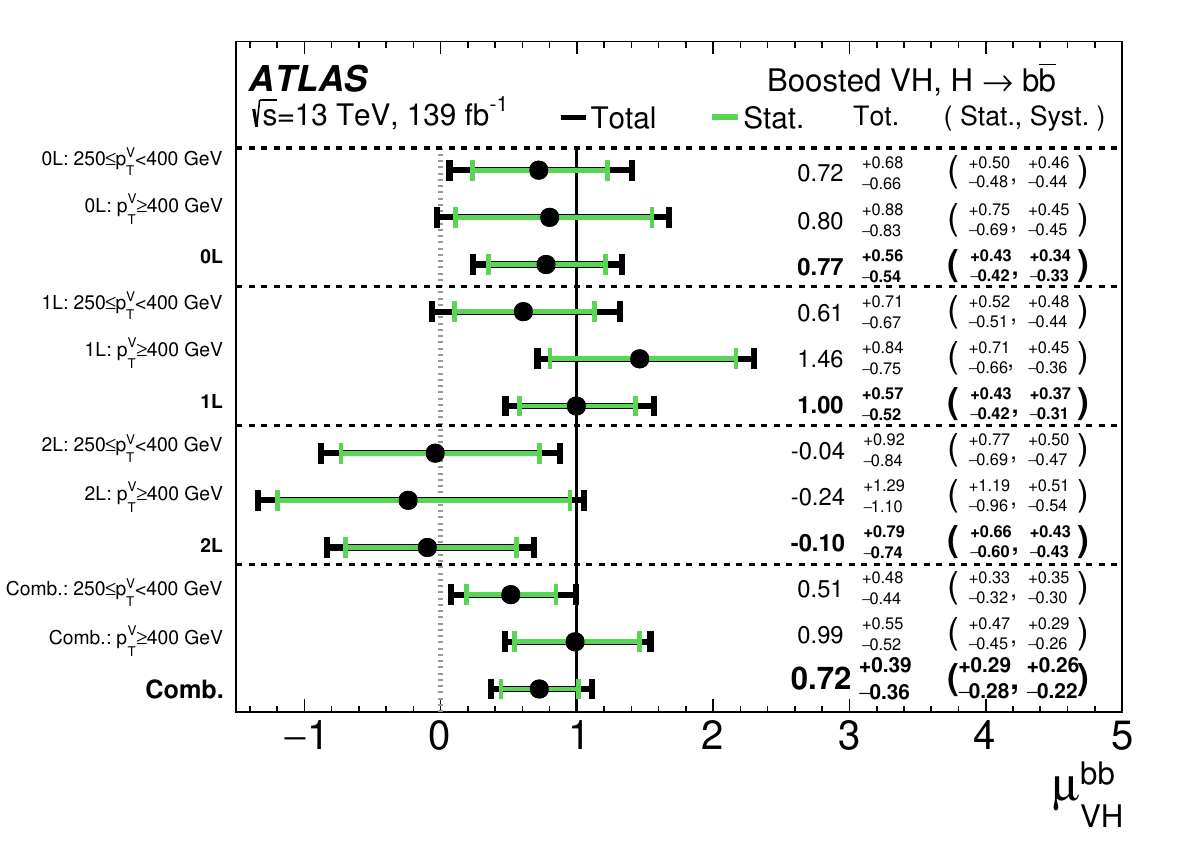}\label{fig:mu_nominaresults}}
\end{center}
\caption{(a) \mJ distribution in data after subtraction of all backgrounds except for the $WZ$ and $ZZ$ diboson processes. The contributions from all lepton channels and signal regions are summed and weighted by their
respective values of the ratio of fitted Higgs boson signal and
background yields. The expected contribution of the associated $WH$
and $ZH$ production of a SM Higgs boson with $m_H = 125$~\GeV\ is
shown scaled by the measured combined signal strength
($\mu_{VH}^{bb}=0.72$). The diboson contribution is normalised to its
best-fit value of $\mu_{VZ}^{bb}=0.91$. The size of the combined
statistical and systematic uncertainty is indicated by the hatched
band. This error band is computed from a full signal-plus-background
fit including all the systematic uncertainties defined in
Section~\ref{sec:systematics}, except for the $VH/VZ$
experimental and theory uncertainties.  (b) Fitted values of the
Higgs boson signal strength parameter, $\mu_{VH}^{bb}$, for $\mH=125$~\GeV\
for the 0-, 1- and 2-lepton channels in different \ptv\ regions
separately and for various
combinations.  \label{fig:jetmass_mu_nominaresults}}
\end{figure}

Figure~\ref{fig:mu_nominaresults} shows the results of: a fit with six
$VH$ POIs measuring the individual signal strengths in each of the
three channels and \ptv\ bins separately; a three $VH$ POI fit
measuring the combined signal strengths in each channel; a two $VH$
POI fit combining all channels in the two \ptv\ bins separately; and
the overall single $VH$ POI combination.
 
For $VZ$ production the fitted signal strength $\mu_{VZ}^{bb}$ is
 
\begin{eqnarray*}
\mu_{VZ}^{bb} = 0.91 ^{+0.29}_{-0.23} = 0.91 \pm 0.15 \mathrm{(stat.)} ^{+0.24}_{-0.17} \mathrm{(syst.)},
\end{eqnarray*}
 
\noindent in agreement with the SM prediction and the $W^{\pm}Z$
differential cross-section measurement performed by ATLAS at high
transverse momentum of the $Z$ boson ($p_\textrm{T}^Z > 220$~\GeV) in
the fully leptonic channel
($W^{\pm}Z \rightarrow \ell'\nu \ell^+ \ell^-$)~\cite{Aaboud:2019gxl}.
The simultaneous fit tests the performance of the analysis on an
irreducible background, the known $VZ$ production, with a topology
similar to the $VH$ signal.  With all three lepton channels combined,
a significance of 5.4 standard deviations is observed for the $VZ$
process, compared to an expectation of 5.7 standard deviations. The
correlation with the $\mu_{VH}^{bb}$ signal strength is approximately
11\%. The statistical uncertainties amount to approximately 60\% of
the total uncertainty. The dominant source of systematic uncertainty
is the background modelling, which has an impact of approximately 0.16
on the result. The source of systematic uncertainty related to
the \largeR\ jet reconstruction follows closely, with an impact of
approximately 0.09 on $\mu_{VZ}^{bb}$.

The cross-sections in the STXS framework are measured separately for
$ZH$ and $WH$ production in two \ptv\ regions,
$250\,\GeV<\ptv<400\,\GeV$ and $\ptv \geq 400\,\GeV$. The analysis
closely follows the strategy used in Ref.~\cite{HIGG-2018-50}.  The
expected signal distributions and acceptance times efficiencies for
each STXS region are estimated from the simulated signal samples by
selecting events using the generator's `truth' information, in
particular the `truth' \ptv, denoted by $p_\textrm{T}^{V,t}$.  The
likelihood function used is different from the one used to extract
the $\mu_{VH}^{bb}$ and $\mu_{VZ}^{bb}$ results presented before.
It has multiple POIs corresponding to the
cross-sections in the four regions used in the analysis, multiplied by
the $H\rightarrow b\bar{b}$ and $V\rightarrow$~leptons branching
fractions. These four regions, i.e. $ZH$ and $WH$ production and the
two $p_\textrm{T}^{V,t}$ bins, are known as reduced stage-1.2 regions in the STXS
framework~\cite{Berger:2019wnu}. The sources of systematic uncertainty are identical to
those defined in Section~\ref{sec:systematics}, except for the theoretical
cross-section and branching fraction uncertainties, which are not included
in the likelihood function because they affect the signal strength
measurements but not the STXS measurements.
 
The cross-sections are not constrained to be positive in the fit. The
measured reduced stage-1.2 $VH$ cross-section times branching fraction
$\sigma \times B$ in each STXS bin, together with the SM predictions
are summarised in Figure~\ref{fig:STXSBinXSPlot} where the red error
bands correspond to the theoretical uncertainty of the fiducial
cross-section prediction in each bin.
The measurements are also reported in Table~\ref{tab:STXSresults} and
are in agreement with the SM predictions from the signal MC sample.
The principal sources of
systematic uncertainties are similar to those affecting
$\mu_{VH}^{bb}$.

These results complement and extend those obtained by the \smallR\ jets
analysis~\cite{CompanionPaper} using the same dataset. The latter
provides a more precise measurement of the cross-section in the
inclusive \ptv\ $>$ 250~\GeV\ region. This can be
attributed
to the larger acceptance at lower $\ptv$ value, the usage of more
precise physics objects calibration and to the use of multivariate
analysis techniques. The results obtained by the two analyses in this
region are compatible within one standard deviation.

\begin{figure}[!htbp]
\centering
\includegraphics[width=0.54\textwidth]{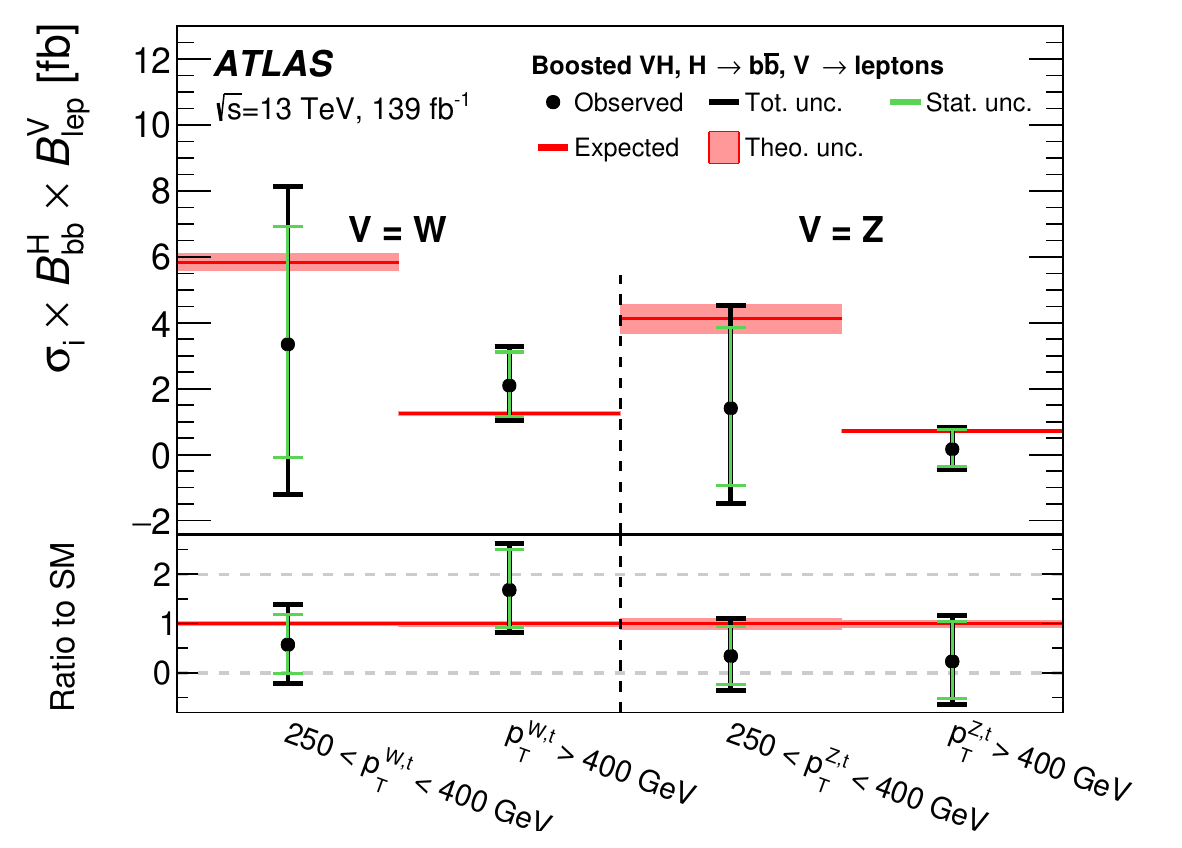}
\caption{Measured
$VH$ reduced stage-1.2 simplified
template cross-sections times the $H \rightarrow \bbbar$ and
$V \rightarrow$~leptons branching fractions.\label{fig:STXSBinXSPlot}}
 
\end{figure}
 
 
\begin{table}[]
 
\caption{Measured and predicted $VH, V \rightarrow \text{leptons}$ reduced stage-1.2 simplified template cross sections times the $H \rightarrow \bbbar$ and
$V \rightarrow$~leptons branching fractions with corresponding uncertainties. All possible $Z$ decays into neutral and charged leptons are considered.}
\label{tab:STXSresults}
 
\begin{center}
 
\begin{tabular}{l|r|rrrr}
\toprule
STXS region $(|y_H| < 2.5, H \rightarrow b\bar{b})$ & SM prediction [fb]  & Result  & (Tot.) & (Stat.) & (Syst.) [fb]  \\
\toprule
$W \rightarrow \ell \nu;~ p_\textrm{T}^{W,t} \in [250, 400]\,$GeV & 5.83 $\pm$ 0.26~~~~ & 3.3 &$^{+4.8}_{-4.6}$ & $ ^{+3.6}_{-3.4}$ & $^{+3.2}_{-3.0} $~~~~~~~ \\  \hline
$W \rightarrow \ell \nu;~ p_\textrm{T}^{W,t} \in [400, \infty]\,$GeV & 1.25 $\pm$ 0.06~~~~ & 2.1 &$^{+1.2}_{-1.1}$ & $ ^{+1.0}_{-0.9}$ & $^{+0.6}_{-0.5} $~~~~~~~ \\ \hline
$Z \rightarrow \ell \ell, \nu\nu;~ p_\textrm{T}^{Z,t} \in [250, 400]\,$GeV & 4.12 $\pm$ 0.45~~~~ & 1.4 &$^{+3.1}_{-2.9}$ & $^{+2.4}_{-2.3}$ & $^{+1.9}_{-1.7}$~~~~~~~ \\ \hline
$Z \rightarrow \ell \ell, \nu\nu;~ p_\textrm{T}^{Z,t} \in [400, \infty]\,$GeV & 0.72 $\pm$ 0.05~~~~ & 0.2 &$^{+0.7}_{-0.6}$ & $^{+0.6}_{-0.5}$ & $ ^{+0.3}_{-0.3}$~~~~~~~ \\
\bottomrule
\end{tabular}
\end{center}
\end{table}

\section{Constraints on anomalous Higgs boson interactions}
\label{sec:eft}

The STXS results presented in Section~\ref{sec:results} are
interpreted in an effective field theory approach where the scale of
new physics is significantly larger than the SM electroweak scale so
as to affect the measured observables at the LHC only through
effective interactions among SM particles.
 
In this SMEFT approach, the SM Lagrangian is
extended with higher-dimensional operators that capture the
low-energy limit effects of a fundamental ultraviolet theory, without a priori
knowledge of this theory~\cite{Contino:2013kra}
 
\begin{equation*}
\label{eq:EFT}
\mathcal{L}_{\textrm{SMEFT}} = \mathcal{L}_{\textrm{SM}} + \sum_d \frac{1}{\Lambda^{d-4}} \left (\sum_i c_i^{(d)} \mathcal{O}_i^{(d)} \right ) ,
\end{equation*}
 
\noindent where $\mathcal{O}_i^{(d)}$ are dimension-$d$ operators and $c_i^{(d)}$ are the corresponding numerical coefficients called Wilson coefficients. All Wilson coefficients are assumed real. In the SM, all Wilson
coefficients are zero. The scale of new physics $\Lambda$ is a free
parameter set to 1~\TeV. In this analysis, the Warsaw basis~\cite{Grzadkowski:2010es} of dimension $d=6$ operators is used, taking into account only the lepton- and baryon-number-conserving ones. Furthermore, it only considers the CP-even terms respecting a U$(3)^5$ flavour symmetry, which affect the $pp \rightarrow V
(\rightarrow \textnormal{leptons}) H (\rightarrow b\overline{b})$
process~\cite{Brivio:2017btx}. The
operators affecting the signal processes are listed in
Table~\ref{tab:EFToperators}~\cite{ATL-PHYS-PUB-2019-042}.

\begin{table}
\centering
\caption{Wilson coefficients and their corresponding dimension-6 operators in the Warsaw formulation considered in this analysis~\cite{Grzadkowski:2010es, ATL-PHYS-PUB-2019-042}.}
\label{tab:EFToperators}
\begin{tabular}{c|c}
\toprule
Coefficient             &  Operator \\
\toprule
 
$c_{H\Box}$        & $(H^{\dagger} H) \Box (H^{\dagger} H)$ \\
$c_{HDD}$          & $(H^{\dagger} D^{\mu} H)^* (H^{\dagger} D_{\mu}H)$ \\
 
$c_{dH}$ 	& $(H^{\dagger} H)(\overline{q}_{p}d_{r}H)$ \\
 
$c_{HW}$      & $H^{\dagger} H  W_{\mu \nu}^I W^{I\mu\nu}$ \\
$c_{HB}$      & $H^{\dagger} H  B_{\mu \nu} B^{\mu\nu}$ \\
$c_{HWB}$     & $H^{\dagger} \tau^I H  W_{\mu \nu}^I B^{\mu\nu}$ \\ \hline
 
$c_{Hl}^{(1)}$      & $ H^{\dagger} i \overleftrightarrow{D}_{\mu} H (\overline{l}_p \gamma^{\mu} l_r)$ \\
$c_{Hl}^{(3)}$      & $H^{\dagger} i \overleftrightarrow{D}_{\mu}^I H (\overline{l}_p \tau^I \gamma^{\mu} l_r)$ \\
 
$c_{He}^{(1)}$      & $H^{\dagger} i \overleftrightarrow{D}_{\mu} H (\overline{e}_p \gamma^{\mu} e_r)$ \\
 
$c_{Hq}^{(1)}$      & $H^{\dagger} i \overleftrightarrow{D}_{\mu} H (\overline{q}_p \gamma^{\mu} q_r)$ \\
$c_{Hq}^{(3)}$      & $H^{\dagger} i \overleftrightarrow{D}_{\mu}^I H (\overline{q}_p \tau^I \gamma^{\mu} q_r)$ \\
$c_{Hu}$           & $H^{\dagger} i \overleftrightarrow{D}_{\mu} H (\overline{u}_p \gamma^{\mu} u_r)$ \\
$c_{Hd}$           & $H^{\dagger} i \overleftrightarrow{D}_{\mu} H (\overline{d}_p \gamma^{\mu} d_r)$ \\ \hline
 
$c_{ll}^{'}$           & $(\overline{l}_p \gamma_{\mu} l_r)(\overline{l}_s \gamma^{\mu} l_t)$ \\
 
\bottomrule
 
\end{tabular}
\end{table}
 
The Wilson coefficients are used to parameterise the STXS and the
Higgs boson decay rates~\cite{ATL-PHYS-PUB-2019-042} from
leading-order predictions~\cite{Brivio:2017btx} and can be constrained
using the STXS measurements presented in
Section~\ref{sec:results}. The parameterisation of the STXS takes into
account the linear terms originating from the interference between SM
and non-SM amplitudes as well as the quadratic ones from the squared
non-SM amplitudes. The former are of order $1/\Lambda^2$ and the
latter of order $1/\Lambda^4$. Given that the current parameterisation
takes neither next-to-leading-order effects nor the interference
between SM and dimension-8 operators into account, the $1/\Lambda^4$
terms are incomplete. Where applicable, fit results will be shown for
both the linear-only parameterisation and the case where quadratic
terms are also included. Since the $gg \rightarrow ZH$ production
cross-section is higher order in QCD, it is kept fixed to its SM
expectation.  The impact of the Wilson coefficients on the
experimental analysis acceptance is not accounted for in this study.
It was, however, verified that the impact was less than 20\% of the
SMEFT parameterization in the kinematic range considered in this study.

Due to the limited number of STXS bins, not all Wilson coefficients
can be measured simultaneously. The interpretations in the SMEFT
framework are carried out with two different approaches. In the first
the four Wilson coefficients the analysis is most sensitive to (after
removing degeneracies) in addition to the operator which affects the
$H\rightarrow b\bar{b}$ decay are varied individually while the others
are kept equal to their SM value of 0. In the second approach, a
principal component decomposition is performed which provides a set of
four linear combinations (corresponding to the number of STXS
measurements) of all the Wilson coefficients of
Table~\ref{tab:EFToperators} and the relative difference of the
branching fraction of the Higgs boson into a pair of $b$-quarks with its
Standard Model prediction ($\mathcal{I}_{BR}$). These combinations of
Wilson coefficients, referred to as eigenvector combinations, are
fitted simultaneously and their uncertainties are uncorrelated. The
combinations are ordered in terms of experimental sensitivity.
 
In the first approach, the four leading Wilson coefficients are
$c_{Hq}^{(3)}$, $c_{Hu}$, $c_{HW}$, and $c_{HWB}$. Operators with
similar or degenerate effects were removed to get this list of
coefficients. The operators $\mathcal{O}_{Hq}^{(1)}$ and
$\mathcal{O}_{Hd}$ impact the analysis similarly to
$\mathcal{O}_{Hu}$. Only $c_{Hu}$ is therefore reported and can be
considered as representative of this ensemble of Wilson
coefficients. The $\mathcal{O}_{dH}$ operator was added to the list
because it modifies the $H\rightarrow b\bar{b}$ vertex, which is
specific to this analysis. The result of the fit is shown in
Figure~\ref{fig:eft_resultsL}, where both the linear-only and
linear-and-quadratic parameterisations are shown. In the case of the
$c_{Hq}^{(3)}$ coefficient the linear-and-quadratic fit slightly
favours a negative value of $c_{Hq}^{(3)}$, but is also compatible
with the SM value of 0 at the 68\%~confidence level (CL).  The
negative-log-likelihood one-dimensional profiles for the individual
fits of the five aforementioned Wilson coefficients are shown in the
Appendix.

\begin{figure}[ht!]
\begin{center}
\includegraphics[width=0.5\textwidth]{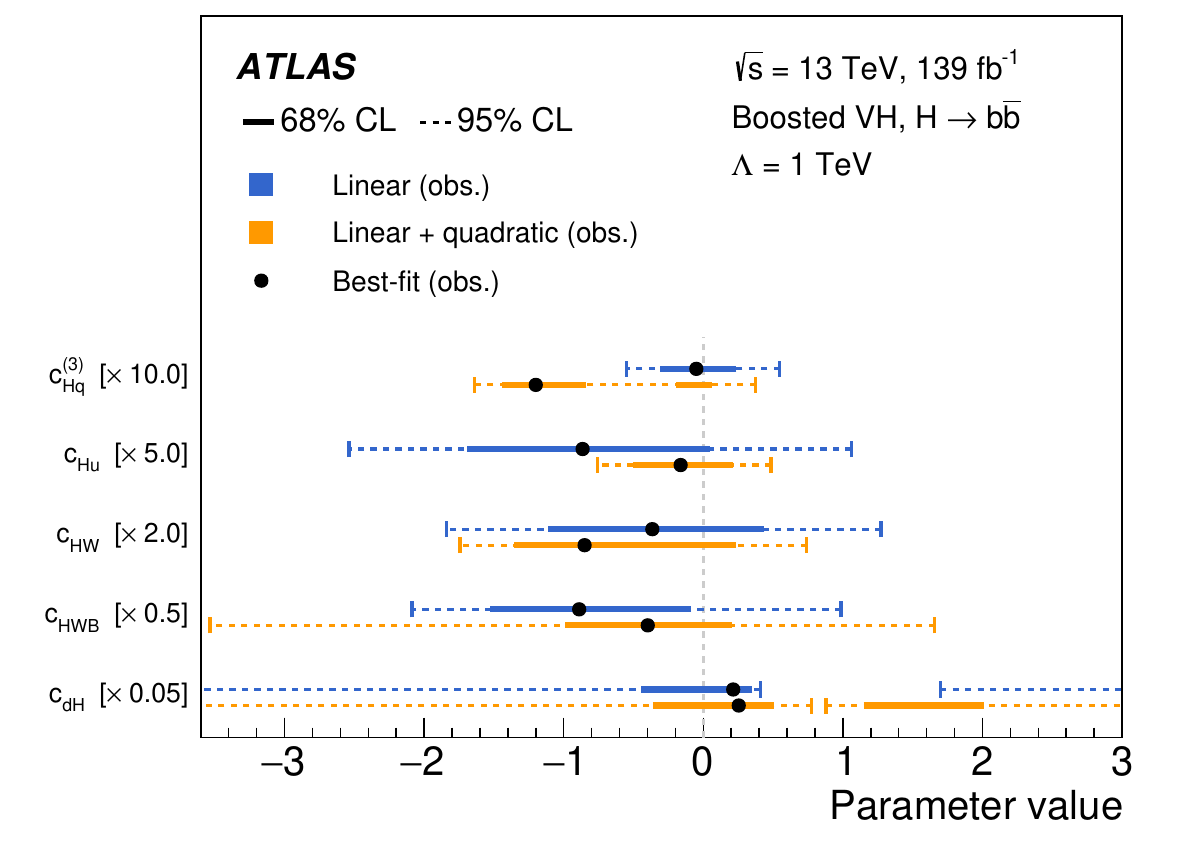}
\label{fig:eft_summary}
\end{center}
\caption{Summary of the observed individual confidence interval at 68\% (solid lines) and 95\% (dashed lines) CL for the $c_{Hq}^{(3)}$, $c_{Hu}$, $c_{HW}$, $c_{HWB}$ and $c_{dH}$ Wilson coefficients from a fit of the STXS, using a linear-only parameterisation (in blue) and including quadratic terms (in orange). \label{fig:eft_resultsL}}
\end{figure}
 
The second approach aims at identifying the Wilson coefficients to
which the measurements are most sensitive (those to which the
measurements are not sensitive manifest themselves as flat directions
in the likelihood) and determining a set of orthogonal linear
combinations of Wilson coefficients.
Only linear terms are considered
and four coefficients denoted $c_{EA}$, $c_{EB}$, $c_{EC}$, and
$c_{ED}$, corresponding to linear combinations of the Wilson
coefficients according to the eigenvectors of the principal component
decomposition, ordered in terms of their experimental sensitivity are
extracted from the data. The parameterisation in terms of the main
Wilson coefficients (reported in Table~\ref{tab:EFToperators}) and of
the branching fraction of the Higgs boson to $b$-quarks, is given in
Table~\ref{tab:TruncatedEV} (coefficients less than 0.10 have been
omitted for better readability).
In contrast to the first approach where the STXS and the partial and
total Higgs decay widths were parameterised independently, this second
approach considers the full branching fraction $H\rightarrow bb$ as
one linear parameter to remove redundancies in operators that only
affect the total Higgs width. The leading combination $c_{EA}$ is
dominated by the $c_{Hq}^{(3)}$ Wilson coefficient as expected from
the fact that it is also the most constrained from the individual
fits. The next-to-leading $c_{EB}$ combination is dominated by the
$c_{Hq}^{(1)}$, $c_{Hu}$, and $c_{Hd}$ Wilson coefficients.  The
$c_{EC}$ and $c_{ED}$ combinations have a larger impact from the
$c_{HW}$ and $c_{HWB}$ coefficients, respectively, with a
non-negligible sensitivity to $c_{Hq}^{(1)}$. The impact of the
variations corresponding to the measurement of each eigenvector
combination on the expected cross-sections in the reduced STXS
measurement bins is shown in Figure~\ref{fig:EFT-eigenvectors}. The
results of the simultaneous fit is given in
Table~\ref{tab:EVResults}. No significant deviation from their
expected SM value is observed.

\begin{table}[]
\centering \caption{Linear combinations of Wilson coefficients
corresponding to the principal component decomposition
eigenvectors (coefficients less than 0.10 have been omitted for
better readability). The corresponding eigenvalues, representing
in the Gaussian approximation the squared inverse uncertainty of
the measured eigenvector, is also indicated.}
 
\label{tab:TruncatedEV}
\begin{tabular}{c | r | p{8cm} }
\toprule
Coefficient & Eigenvalue & Eigenvector combination \\
\toprule
$c_{EA}$ & 1500.0 & $ 0.99 \cdot c_{Hq}^{(3)} +0.11 \cdot c_{Hu}     $ \\
\hline
$c_{EB}$ & 26.9   & $ 0.82 \cdot c_{Hu}       -0.49 \cdot c_{Hq}^{(1)} -0.24 \cdot c_{Hd}  -0.13 \cdot c_{Hq}^{(3)}  $   \\
\hline
$c_{EC}$ & 2.2    & $ 0.67 \cdot \mathcal{I}_{BR}       +0.66 \cdot c_{HW}       +0.18 \cdot c_{Hq}^{(1)}  -0.16 \cdot c_{Hl}^{(3)} +0.14 \cdot c_{HWB}  +0.12 \cdot c_{ll}^{'}    $  \\
\hline
$c_{ED}$ & 0.1    & $ 0.70 \cdot c_{Hq}^{(1)} +0.52 \cdot c_{HWB}      +0.27 \cdot c_{Hu}      -0.27 \cdot c_{HW} -0.24 \cdot c_{Hd}    + 0.13 \cdot c_{HB}   $ \\
\bottomrule
\end{tabular}
\end{table}
 
\begin{figure}[!htbp]
\centering
\includegraphics[width=0.5\textwidth]{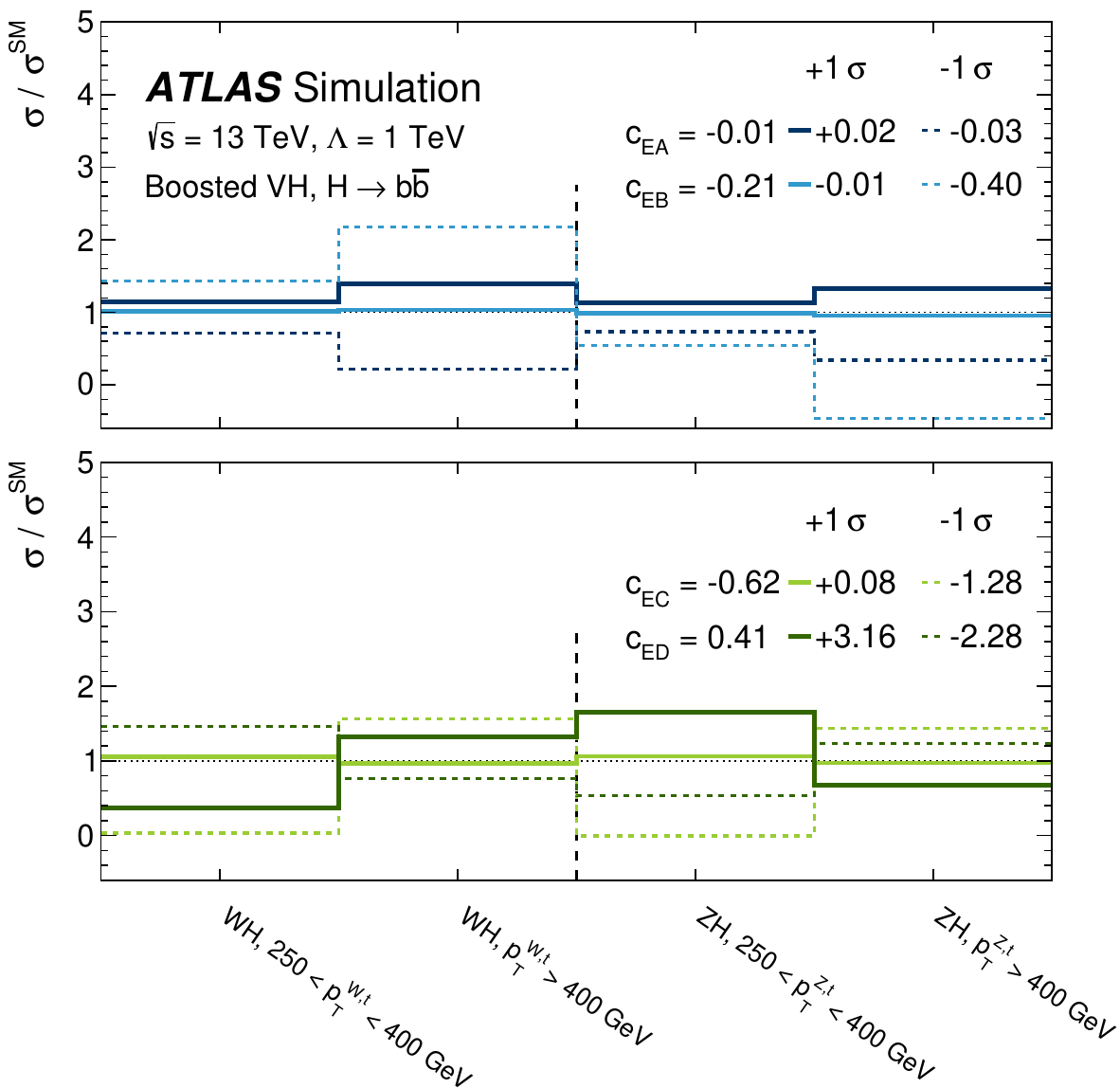}
\caption{The impact of the measured $+1\sigma$ (solid) and $-1\sigma$ (dashed)  variations of the four eigenvectors on the reduced STXS bins.\label{fig:EFT-eigenvectors}}
\end{figure}
 
\renewcommand{\arraystretch}{1.4}
\begin{table}[]
 
\caption{Expected
and observed best-fit values and associated uncertainties (68\%
CL) from a simultaneous fit of the four coefficients corresponding
to the eigenvector combinations.}
 
\centering \begin{tabular}{c|c|c} \toprule  Coefficient
& Expected & Observed \\ \toprule $c_{EA}$ &
$0.000^{+0.030}_{-0.027}$ & $-0.010^{+0.027}_{-0.025}$\\ $c_{EB}$
& $0.00^{+0.20}_{-0.19}$ & $-0.21^{+0.19}_{-0.20}$ \\ $c_{EC}$ &
$0.00^{+0.71}_{-0.67}$ & $-0.62^{+0.70}_{-0.66}$ \\ $c_{ED}$ &
$0.0^{+2.8}_{-2.7}$ &
~~$0.4^{+2.8}_{-2.7}$ \\ \bottomrule \end{tabular}   \label{tab:EVResults}
\end{table}
\renewcommand{\arraystretch}{1.0}

 
\FloatBarrier

\section{Conclusion}
\label{sec:conclusion}
Measurements of a Higgs boson decaying into a $b$-quark pair, produced
in association with a vector boson at high transverse momentum
subsequently decaying into a pair of leptons (electrons, muons and/or
neutrinos) are performed. The results are based on the
Run~2 dataset of $pp$ collision data collected at \sqsthirt\ by the ATLAS detector at the LHC,
corresponding to an integrated luminosity of 139~fb$^{-1}$. These are the most precise measurements
currently available in the high transverse momentum regime for this process. The Higgs
boson is reconstructed from a single \largeR\ jet to enhance the
sensitivity in the high-\pt\ regime.

The significance of the measurement of the SM $VH$ process is 2.1 standard deviations. The expected significance corresponding to the production of the SM $VH$ process is 2.7 standard deviations. For the diboson process ($VZ$), which is measured simultaneously,  the observed significance is 5.4 standard deviations, in good agreement with the expected diboson significance of 5.7 standard deviations.
The cross-sections for the associated production of a
Higgs boson decaying into a $b$-quark pair with an electroweak gauge
boson $W$ or $Z$ decaying into leptons are measured in the simplified
template cross-section framework in two \ptv\ regions:
$250\,\GeV<\ptv<400\,\GeV$ and $\ptv \geq 400\,\GeV$. All results are
in agreement with SM predictions and are interpreted in terms of
constraints on anomalous couplings in the framework of a Standard
Model effective field theory.

 
\clearpage
\section*{Appendix}
\label{appendix}
 
This appendix contains the negative-log-likelihood profile for the
individual fits of the $c_{Hq}^{(3)}$, $c_{Hu}$, $c_{HW}$, $c_{HWB}$
and $c_{dH}$ Wilson coefficients. In each fit all coefficients other
than the one individually varied are set to zero. Each fit is
performed twice, once with and once without the quadratic terms taken
into account, as described in Section~\ref{sec:eft}.
 
\begin{figure}[tbph]
\centering
\includegraphics[width=0.33\linewidth]{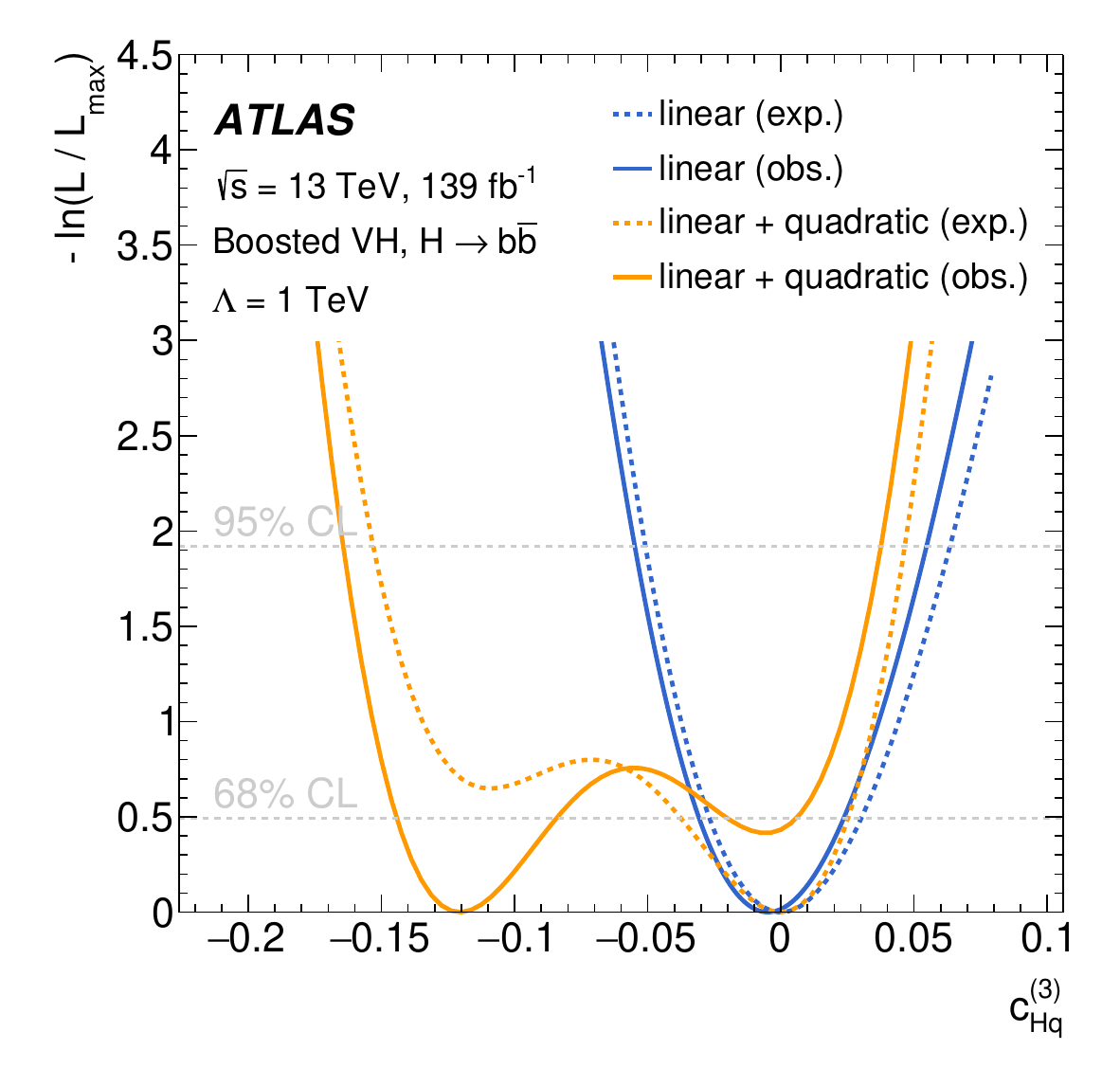}
\includegraphics[width=0.33\linewidth]{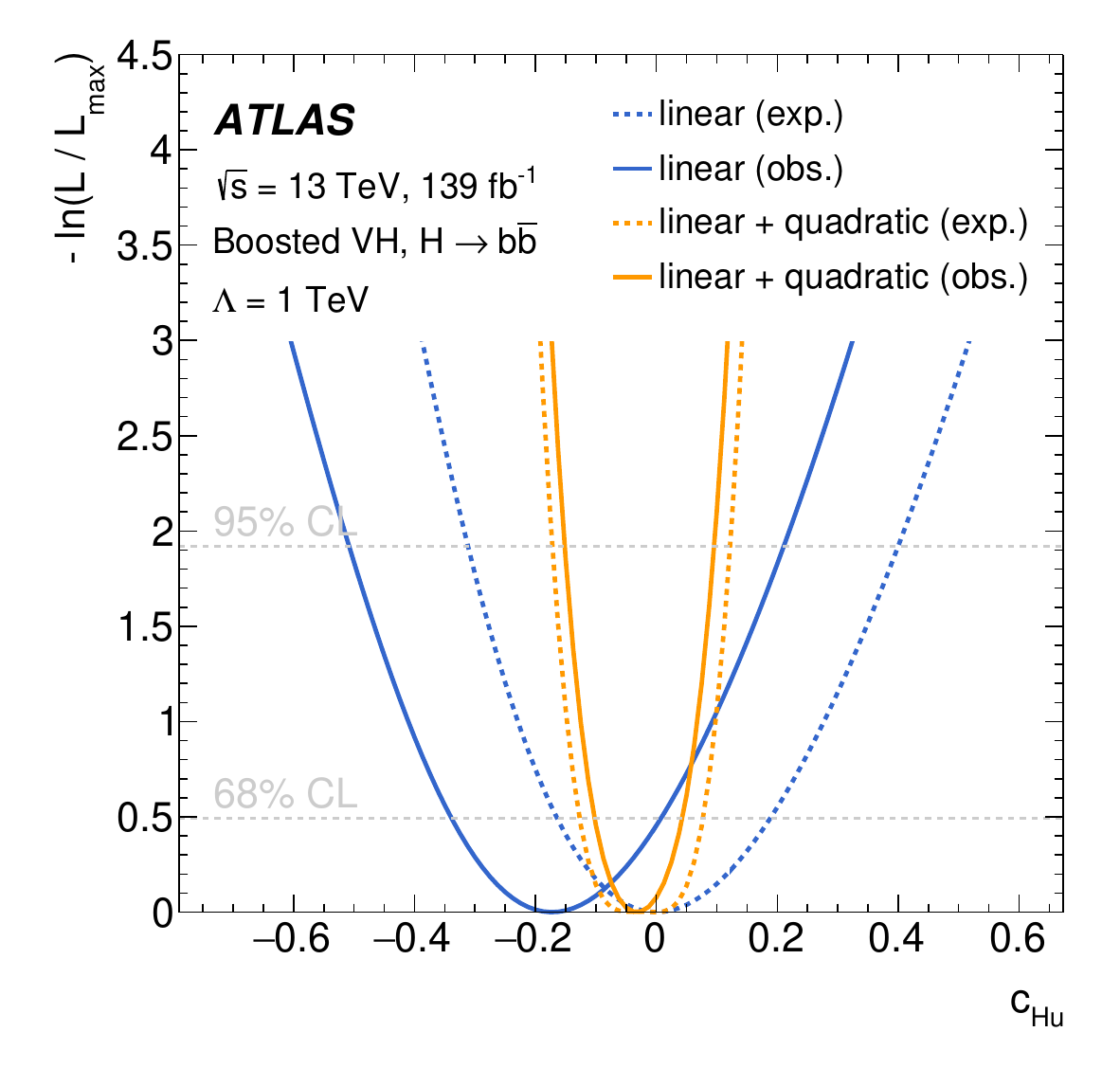}
\includegraphics[width=0.33\linewidth]{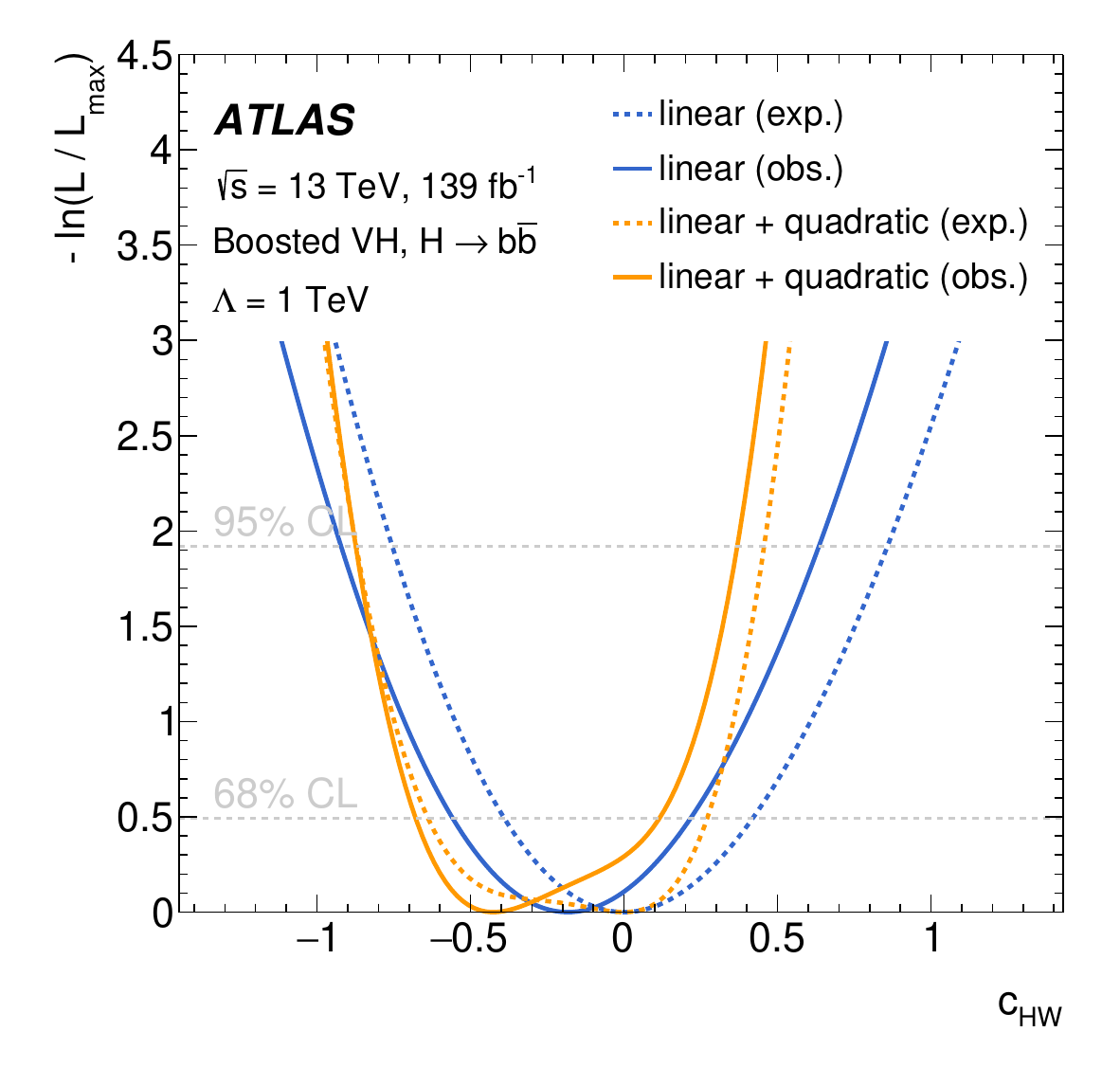}
\includegraphics[width=0.33\linewidth]{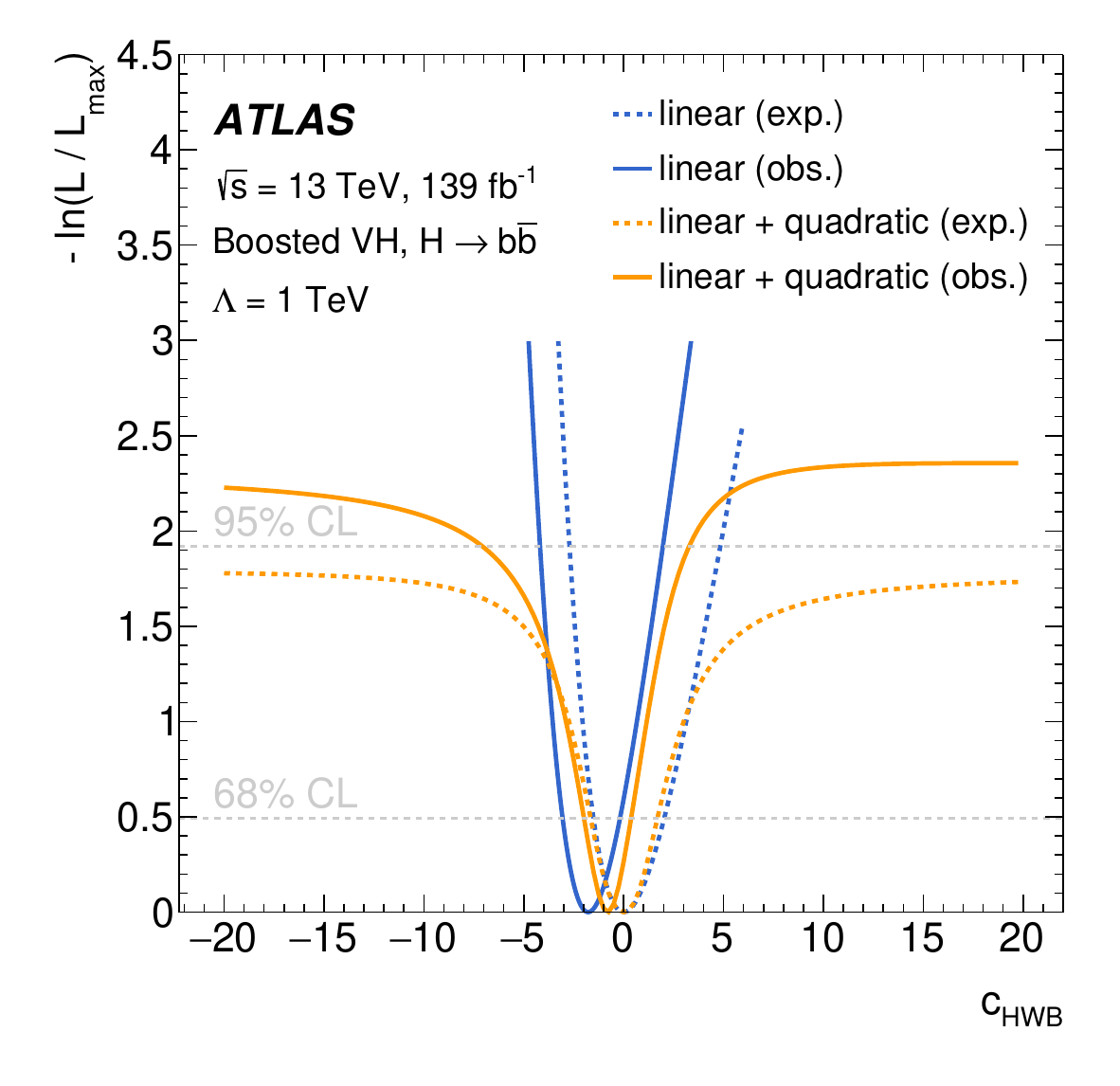}
\includegraphics[width=0.33\linewidth]{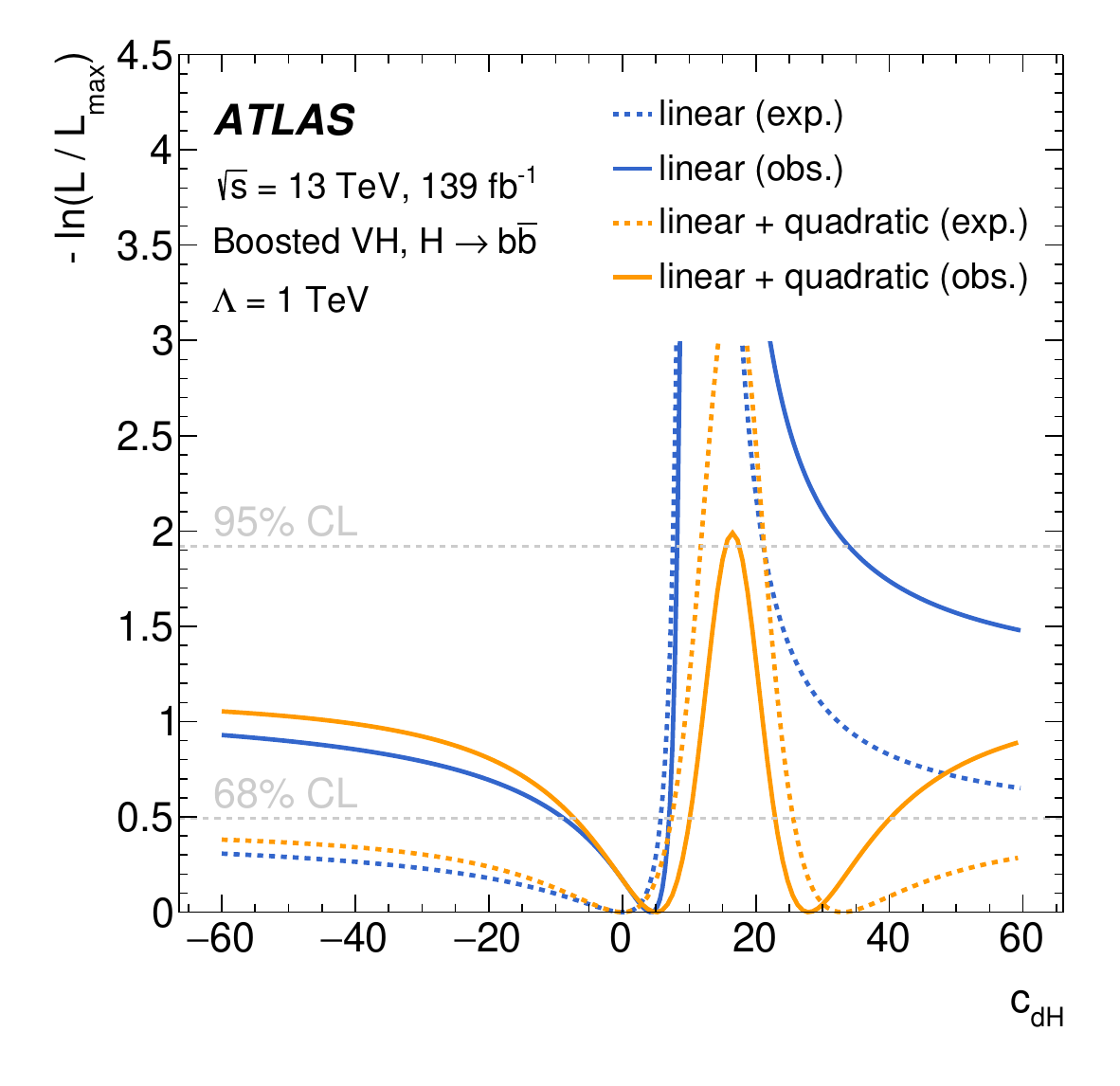}
\caption{The negative log-likelihood profile as a function of the
variations of single Wilson coefficients $c_{Hq}^{(3)}$, $c_{Hu}$,
$c_{HW}$, $c_{HWB}$ and $c_{dH}$. Wilson coefficients of all
other operators are set to zero. The fits are performed using a
linear-only parametrisation of the $VH$ production cross section
and the Higgs decay branching fraction (blue lines), and using a
parametrisation that includes both linear and quadratic terms into
account (orange lines). Observed results are drawn in thick solid
lines, thin dashed lines indicate expected results. The 68\% and
95\% CL lines are also indicated.
\label{fig:1d_scans_production_decay}}
\end{figure}

\clearpage

\section*{Acknowledgements}
 

We thank CERN for the very successful operation of the LHC, as well as the
support staff from our institutions without whom ATLAS could not be
operated efficiently.
 
We acknowledge the support of ANPCyT, Argentina; YerPhI, Armenia; ARC, Australia; BMWFW and FWF, Austria; ANAS, Azerbaijan; SSTC, Belarus; CNPq and FAPESP, Brazil; NSERC, NRC and CFI, Canada; CERN; ANID, Chile; CAS, MOST and NSFC, China; COLCIENCIAS, Colombia; MSMT CR, MPO CR and VSC CR, Czech Republic; DNRF and DNSRC, Denmark; IN2P3-CNRS and CEA-DRF/IRFU, France; SRNSFG, Georgia; BMBF, HGF and MPG, Germany; GSRT, Greece; RGC and Hong Kong SAR, China; ISF and Benoziyo Center, Israel; INFN, Italy; MEXT and JSPS, Japan; CNRST, Morocco; NWO, Netherlands; RCN, Norway; MNiSW and NCN, Poland; FCT, Portugal; MNE/IFA, Romania; JINR; MES of Russia and NRC KI, Russian Federation; MESTD, Serbia; MSSR, Slovakia; ARRS and MIZ\v{S}, Slovenia; DST/NRF, South Africa; MICINN, Spain; SRC and Wallenberg Foundation, Sweden; SERI, SNSF and Cantons of Bern and Geneva, Switzerland; MOST, Taiwan; TAEK, Turkey; STFC, United Kingdom; DOE and NSF, United States of America. In addition, individual groups and members have received support from BCKDF, CANARIE, Compute Canada, CRC and IVADO, Canada; Beijing Municipal Science \& Technology Commission, China; COST, ERC, ERDF, Horizon 2020 and Marie Sk{\l}odowska-Curie Actions, European Union; Investissements d'Avenir Labex, Investissements d'Avenir Idex and ANR, France; DFG and AvH Foundation, Germany; Herakleitos, Thales and Aristeia programmes co-financed by EU-ESF and the Greek NSRF, Greece; BSF-NSF and GIF, Israel; La Caixa Banking Foundation, CERCA Programme Generalitat de Catalunya and PROMETEO and GenT Programmes Generalitat Valenciana, Spain; G\"{o}ran Gustafssons Stiftelse, Sweden; The Royal Society and Leverhulme Trust, United Kingdom.
 
The crucial computing support from all WLCG partners is acknowledged gratefully, in particular from CERN, the ATLAS Tier-1 facilities at TRIUMF (Canada), NDGF (Denmark, Norway, Sweden), CC-IN2P3 (France), KIT/GridKA (Germany), INFN-CNAF (Italy), NL-T1 (Netherlands), PIC (Spain), ASGC (Taiwan), RAL (UK) and BNL (USA), the Tier-2 facilities worldwide and large non-WLCG resource providers. Major contributors of computing resources are listed in Ref.~\cite{ATL-SOFT-PUB-2020-001}.
 

\printbibliography

\clearpage
 
 
\begin{flushleft}
\hypersetup{urlcolor=black}
{\Large The ATLAS Collaboration}

\bigskip

\AtlasOrcid[0000-0002-6665-4934]{G.~Aad}$^\textrm{\scriptsize 102}$,    
\AtlasOrcid[0000-0002-5888-2734]{B.~Abbott}$^\textrm{\scriptsize 128}$,    
\AtlasOrcid[0000-0002-7248-3203]{D.C.~Abbott}$^\textrm{\scriptsize 103}$,    
\AtlasOrcid[0000-0002-2788-3822]{A.~Abed~Abud}$^\textrm{\scriptsize 36}$,    
\AtlasOrcid[0000-0002-1002-1652]{K.~Abeling}$^\textrm{\scriptsize 53}$,    
\AtlasOrcid[0000-0002-2987-4006]{D.K.~Abhayasinghe}$^\textrm{\scriptsize 94}$,    
\AtlasOrcid[0000-0002-8496-9294]{S.H.~Abidi}$^\textrm{\scriptsize 167}$,    
\AtlasOrcid[0000-0002-8279-9324]{O.S.~AbouZeid}$^\textrm{\scriptsize 40}$,    
\AtlasOrcid{N.L.~Abraham}$^\textrm{\scriptsize 156}$,    
\AtlasOrcid[0000-0001-5329-6640]{H.~Abramowicz}$^\textrm{\scriptsize 161}$
% ...
\bigskip
\\

% ...
$^{36}$CERN, Geneva; Switzerland.\\
% ...
$^{40}$Niels Bohr Institute, University of Copenhagen, Copenhagen; Denmark.\\
% ...
$^{53}$II. Physikalisches Institut, Georg-August-Universit\"{a}t G\"ottingen, G\"ottingen; Germany.\\
% ...
$^{94}$Department of Physics, Royal Holloway University of London, Egham; United Kingdom.\\
% ...
$^{102}$CPPM, Aix-Marseille Universit\'e, CNRS/IN2P3, Marseille; France.\\
$^{103}$Department of Physics, University of Massachusetts, Amherst MA; United States of America.\\
% ...
$^{128}$Homer L. Dodge Department of Physics and Astronomy, University of Oklahoma, Norman OK; United States of America.\\
% ...
$^{156}$Department of Physics and Astronomy, University of Sussex, Brighton; United Kingdom.\\
% ...
$^{161}$Raymond and Beverly Sackler School of Physics and Astronomy, Tel Aviv University, Tel Aviv; Israel.\\
% ...
$^{167}$Department of Physics, University of Toronto, Toronto ON; Canada.\\
% ...

\end{flushleft}


\end{document}